\documentclass[letterpaper]{JHEP3}
\usepackage[latin9]{inputenc}
\usepackage{amsmath}
\usepackage{amssymb}

\makeatletter

\pdfpageheight\paperheight
\pdfpagewidth\paperwidth

\pdfoutput=1

\usepackage{amsfonts}\usepackage{cite}

\def\be{\begin{equation}}
\def\ee{\end{equation}}
\def\bea{\begin{eqnarray}}
\def\eea{\end{eqnarray}}

\title{Null infinity and extremal horizons in AdS-CFT}

\author{Andrew Hickling\\
{\it Theoretical Physics Group, Blackett Laboratory, Imperial College, London SW7 2AZ, U.K. } \\
\email{a.hickling12@imperial.ac.uk}
}

\author{James Lucietti\\
{\it School of Mathematics and Maxwell Institute for Mathematical Sciences, University of  Edinburgh, King's Buildings, Edinburgh, EH9 3JZ, U.K.} \\
\email{j.lucietti@ed.ac.uk}
}

\author{Toby Wiseman\\
{\it Theoretical Physics Group, Blackett Laboratory, Imperial College, London SW7 2AZ, U.K. } \\
\email{t.wiseman@imperial.ac.uk}
}

\date{today}
\abstract{

We consider AdS gravity duals to CFT on background spacetimes with a null infinity. Null infinity on the conformal boundary may extend to an extremal horizon in the bulk. For example it does so for Poincar\'e-AdS, although does not for planar Schwarzschild-AdS. If null infinity does extend into an extremal horizon in the bulk, we show that the bulk near-horizon geometry is determined by the geometry of the boundary null infinity. Hence the `infra-red' geometry of the bulk is fixed by the large scale behaviour of the CFT spacetime. 
In addition the boundary stress tensor must have a particular  decay at null infinity.
As an application, we argue that for CFT on asymptotically flat backgrounds, any static bulk dual containing an extremal horizon extending from the boundary null infinity, must have the near-horizon geometry of Poincar\'e-AdS.  We also discuss a class of boundary null infinity that cannot extend to a bulk extremal horizon, although we give evidence that they can extend to an analogous null surface in the bulk which possesses an associated scale-invariant `near-geometry'. 

}

\makeatother

\begin{document}

\section{Introduction}

\label{sec:intro}

The AdS-CFT duality \cite{Maldacena:1997re} is a truly remarkable tool that allows the study of certain strongly coupled field theory phenomena in terms of a dual gravitational problem. Increasingly, such problems involve the study of bulk gravity in rather exotic contexts which have not previously been considered. In particular, this has opened up the study of CFT in situations which have not been explored using conventional field theory methods, such as strongly coupled CFT on {\it curved} backgrounds, see~\cite{Marolf:2013ioa} for a review.  For example, the study of CFT at strong coupling on a black hole background, maps to the study of new types of bulk geometries termed black droplets and black funnels~\cite{Hubeny:2009ru, Hubeny:2009kz}. By now, several examples of such geometries have been constructed numerically~\cite{Figueras:2011va, Santos:2012he, Figueras:2013jja, Fischetti:2013hja, Santos:2014yja}.

One of the important tools that has been introduced in recent years to study gravity in higher dimensional contexts is the extremal horizon and its near-horizon geometry, see~\cite{Kunduri:2013gce} for a review. The goal of this paper is to investigate the role of extremal horizons in the bulk physics of AdS-CFT. Specifically, we will consider extremal horizons which extend to the conformal boundary.

Consider a CFT on a general static spacetime. We would expect the vacuum state to be dual to a static bulk spacetime. Suppose the boundary spacetime has asymptotic regions, such as Minkowski spacetime which is asymptotically flat. These asymptotic regions determine the large scale properties of the spacetime. Now, assuming that the CFT spectrum is not gapped, then the lowest lying excitations will correspond to the long wavelength physics in these asymptotic regions. We expect this will be captured by a dual bulk spacetime ending on an extremal horizon. Extremal horizons have an infinite throat whose geometry -- often called the near-horizon geometry -- is itself also a solution of the Einstein equations. The long wavelength physics in the CFT should be described in terms of perturbations living in this near-horizon region of the bulk, which from the boundary perspective will have small energies due to the large redshift near the horizon.

There are various natural questions one might ask. How does the asymptotic behaviour of the boundary metric correspond to properties of the bulk extremal horizon?  When can a bulk extremal horizon end on the boundary? Is a bulk extremal horizon the most general geometry that describes large scale behaviour? It is precisely these issues that we will examine in this paper. 

The first point we will clarify is that if the boundary spacetime has an asymptotic region that contains a null infinity, as for example for Minkowski spacetime, then a bulk extremal horizon may extend to the boundary and be the continuation of this boundary null infinity. However, for this to happen, the boundary null infinity must itself be conformal to an extremal horizon. We term such a null infinity as being \emph{extremal}. In the case of Poincar\'e-AdS, the extremal horizon is the bulk Cauchy horizon, and indeed extends to null infinity of the boundary Minkowski spacetime which is extremal.

In fact we shall show that for such a bulk extremal horizon that is smooth and has no other boundaries, its near-horizon geometry is precisely determined by the structure of the boundary extremal null infinity. As a consequence, any deformation of the boundary metric that preserves the asymptotic structure will generically leave such bulk extremal horizons unchanged. For example, taking the usual Poincar\'e-AdS case and deforming the Minkowski boundary metric to some other asymptotically flat static geometry -- such as Schwarzschild as for black droplets \cite{Figueras:2011va} -- will generically leave the extremal horizon unperturbed for the vacuum geometry. In the example of droplets the near-horizon geometry would still be that of Poincar\'e-AdS. However, any deformation that changes the structure of infinity on the boundary, for example introducing an asymptotically conical rather than flat geometry, will deform the bulk extremal horizon geometry, in this case away from that of Poincar\'e-AdS.\footnote{CFT with defects arising from vortices have been considered in~\cite{Dias:2013bwa}. In their case it is the matter fields rather than the CFT background metric which possess the defect.}

The second issue we will address is under what conditions may null infinity on the boundary extend to an extremal horizon in the bulk. This appears to be a harder question, and we will only be able to give obstructions to extension. The first obstruction as mentioned above is that the boundary null infinity must be extremal. For example, if we take the boundary to be Minkowski with a spatial direction periodically identified, then boundary null infinity is no longer extremal, and indeed the possible bulk vacuum duals do not have smooth extremal horizons.

The second obstruction we discuss is given by the asymptotic fall-off of the boundary stress tensor. If the boundary stress tensor does not satisfy a particular decay law near null infinity, then null infinity on the boundary cannot extend to a bulk extremal horizon. As an example of this, consider again Poincar\'e-AdS corresponding to the bulk dual of a CFT on Minkowski in the vacuum state. This has zero stress tensor. However, consider instead the static bulk associated to the thermal state, planar-AdS Schwarzschild. This has a constant stress tensor that does not decay at infinity, and correspondingly, the boundary null infinity does not extend to a bulk extremal horizon. Rather the bulk has  a non-extremal horizon, and there is a gap in the theory associated to the thermal mass.

While certain long range behaviour in static CFT states with no gap is determined by the near-horizon geometry of static bulk extremal horizons, we might ask whether all long range behaviour is determined in this way. We consider certain static boundary metrics which have null infinities that are not extremal, but nonetheless exhibit a 
particular asymptotic scaling symmetry, and hence have a closely related form to extremal null infinity. We term these as having a  \emph{twisted} null infinity. Correspondingly we argue that the dual static bulk may end in a geometry analogous to an extremal horizon, in that it has a near scaling limit, and infinite redshift.
However, unlike extremal horizons, we believe these are generically null singularities. 

For ease of applicability we begin the paper with section \ref{sec:summary} that provides a summary of our results and provides explicit examples illustrating these. In the remainder of the paper we provide the arguments for these results. We begin in section \ref{sec:null} by defining the class of null infinities we are concerned with, those that are extremal and those that are twisted. Then in section \ref{sec:obstruction} we argue that the asymptotic stress tensor provides an obstruction to the extension of boundary extremal null infinity to a bulk extremal horizon. In section \ref{sec:extremal} we first review the relevant features of extremal horizons before then showing how a bulk extremal horizon meets the conformal boundary at an extremal null infinity.  Finally in section \ref{sec:twisted} we discuss the boundary twisted null infinities that we believe have vacuum bulk duals ending on a singular null surface of infinite redshift. 

\section{Summary and examples}

\label{sec:summary}

In this section we provide a summary of our main results and give examples that illustrate them in practice. In the remainder of the paper we derive the results described here.\\

\subsection{Setup and main results}

{\bf Setup.} We will consider $d+1$ dimensional spacetimes $(M,g)$, with a conformal boundary, satisfying the vacuum Einstein equations with a negative cosmological constant. We will set the AdS$_{d+1}$ length to one so that our conventions are such that
\be
R_{AB}= - d \, g_{AB} \; .  \label{Einstein}
\ee
We will consider solutions with a conformal boundary representative $(\partial M, h)$ which itself possesses a null infinity $\mathcal{I}$.  We will also introduce an alternate conformal frame $(\overline{\partial M}, \overline{h})$, that is a `conformal compactification' of $(\partial M, h)$, in which $\mathcal{I}$ is a smooth null hypersurface.\footnote{This is a slight abuse of language since we will actually work with non-compact $\overline{\partial M}$.}  We will assume the  `bulk' spacetime $(M,g)$ possesses a Killing field $\xi$, such that the representatives $(\partial M, h)$  and $(\overline{\partial M}, \overline{h})$ inherit this isometry. We will often assume that $\xi$ is hypersurface-orthogonal, which includes static spacetimes ($\xi$ timelike) or certain cosmologies ($\xi$ spacelike). \\

\noindent {\bf Definition}. Null infinity $\mathcal{I}$ is said to be {\it extremal} if there exists a conformal frame $(\overline{\partial M}, \overline{h})$ such that $\mathcal{I}$ is a smooth degenerate Killing horizon. \\

\noindent In section \ref{sec:Enull} we show that necessary and sufficient conditions for null infinity $\mathcal{I}$ to be extremal are that the norm $N = h_{\mu\nu} \xi^\mu \xi^\nu$ and its first derivatives are bounded near $\mathcal{I}$. This includes various notable cases, including static asymptotically flat, or conical, spacetimes, and also certain cosmologies which approach Milne like spacetimes at late times. \\

\noindent {\bf Results}. Consider a spacetime $(M,g)$, with a conformal boundary, that satisfies (\ref{Einstein}). Suppose it contains an extremal horizon $\mathcal{N}$ defined by a hypersurface-orthogonal Killing field $\xi$, with cross-sections $H$, which intersects the conformal boundary. The following results apply:

\begin{enumerate}
\item There exists a conformal frame $(\partial M, h)$ with an extremal null infinity $\mathcal{I}$ such that $\partial \mathcal{N} = {\mathcal{I}}$. Thus, null infinity $\mathcal{I}$ of a boundary spacetime $(\partial M, h)$ may extend to a bulk extremal horizon only if it is an extremal null infinity.

\item The near-horizon geometry of $\mathcal{N}$ possesses a conformal boundary and satisfies the elliptic near-horizon equations on $H$, with the boundary condition fully determined by the geometry of $\mathcal{I}$ (if there are no other boundaries). For a given topology $H$, there may be no solution, a discrete set of  solutions, or a finite dimensional moduli space of solutions.

\item  The boundary stress tensor in the conformal frame $(\partial M, h)$ must have a particular decay near $\mathcal{I}$.  Therefore, if the stress tensor violates this precise decay law,  $\mathcal{I}$ cannot extend to an extremal horizon in the bulk.
\end{enumerate}

\noindent We will prove these results in section \ref{sec:extremal}.  As a concrete application of these results, in section \ref{sec:application} we argue that if $(\partial M, h)$ is a static asymptotically Minkowski spacetime, then any static bulk extremal horizon extending from boundary null infinity must have the  near-horizon geometry of Poincar\'e-AdS. Further, the boundary stress-tensor must decay  as $\mathcal{O}(\rho^{-d})$, where $\rho$ is the radial coordinate in Minkowski spacetime.  \\

\noindent It is of interest to relax the notion of extremal null infinity by dropping the requirement that $\mathcal{I}$ is a smooth null hypersurface in $(\overline{\partial M}, \overline{h})$. As we discuss in section \ref{sec:Tnull} this leads to a more general notion of {\it twisted null infinity} which is conformally related to scale invariant geometries. \\

\noindent
{\bf Conjecture 1.} There exist static bulk spacetimes $(M,g)$ that end on a singular null hypersurface $\mathcal{N}$ which intersects the conformal boundary, with the following properties:

\begin{itemize} 
\item There exists a conformal frame $(\partial M, h)$ with a twisted null infinity $\mathcal{I}$ such that $\partial \mathcal{N} = \mathcal{I}$.
\item There is a scale-invariant near-geometry for $\mathcal{N}$, which is determined by a system of elliptic equations more general than those for an extremal horizon, with the boundary condition fully determined by the geometry of $\mathcal{I}$ (if there are no other boundaries).
\end{itemize}

\noindent In section \ref{sec:twisted} we prove two non-existence theorems concerning such scale-invariant geometries and also provide evidence for the above conjecture via  a linearised calculation. 

\subsection{Examples}

\label{sec:examples}

We will now provide illustrative examples of the results concerning extremal horizons. Our examples are all for the case of a static-timelike isometry, although by analytic continuation one can also construct analogous examples with a static-spacelike isometry. Unfortunately we cannot give simple examples for the case of boundary twisted null infinity and their bulk dual, but will discuss evidence for their existence in section \ref{sec:twisted}.

\subsubsection{Poincar\'e-AdS}
\label{sec:poincare}

Consider Poincar\'e-AdS$_{d+1}$,
\begin{equation}
g =\frac{-dt^2 + dx^i dx^i + dz^{2}}{z^{2}}  \; ,
\end{equation}
where $i=1, \dots, d-1$.
As is well known, this is the bulk dual to a CFT on $d$-dimensional Minkowski spacetime in the vacuum state. The conformal boundary is at $z =0$ and its conformal class is that of $d$-dimensional Minkowski spacetime. In spherical polar coordinates
\be
h_{\text{Mink}} = -dt^2+d\rho^2 +\rho^2 d\Omega_{(d-2)}^2  \; ,  \label{mink}
\ee
which of course possesses a null infinity at $\rho \to \infty, t \to \pm \infty$. Conformally compactifying we obtain,
\be
\bar{h}= \rho^{-2} h_{\text{Mink}} = \frac{-dt^2+d\rho^2}{\rho^2} + d\Omega_{(d-2)}^2  \; , \label{eq:boundaryMink}
\ee
which is a near-horizon geometry AdS$_2\times S^{d-2}$ with an extremal horizon at $\rho \to \infty$ corresponding to the null infinity of Minkowski spacetime.

The Poincar\'e horizon in the bulk is an extremal horizon, and is located at $z\rightarrow\infty$
and $x^i \rightarrow\infty$ such that $x^i/z$ is held finite.
We may use coordinates adapted to this extremal horizon~\cite{Figueras:2011va, Kunduri:2013gce},
\begin{equation}
\label{eq:exAdSbulk}
g =\psi^{2}\left( \frac{-dt^{2}+d\rho^2}{\rho^2} \right)+\frac{d\psi^{2}}{\psi^{2}-1}+(\psi^{2}-1) d\Omega_{(d-2)}^{2}  \; ,
\end{equation}
where $\psi \geq 1$. The horizon is now at $\rho \to \infty$ and the conformal boundary is at $\psi \to \infty$. In fact we may now recognise that Poincar\'e-AdS is its own near-horizon geometry. Indeed, performing the usual scaling $(t, \rho) \to (\epsilon^{-1} t, \epsilon^{-1} \, \rho)$ simply leaves this metric invariant. 

We now see a clear illustration of our Result 1 and Result 2. Firstly, the boundary null infinity of Minkowski is extremal (being conformal to the extremal horizon in (\ref{eq:boundaryMink})) and extends into the bulk as the bulk extremal horizon in \eqref{eq:exAdSbulk}. Secondly, the bulk near-horizon geometry (\ref{eq:exAdSbulk}) has a conformal boundary at $\psi \to \infty$, whose geometry $\psi^{-2}g|_{\psi \to \infty}$ is precisely the conformal compactification of the boundary Minkowski null infinity (\ref{eq:boundaryMink}). Finally, note that Result 3 gives no obstruction to extending the boundary null infinity to a bulk horizon, as the boundary stress tensor in this case vanishes identically.

The above example is of course a rather trivial illustration of our results. In fact, we may deduce  a much more non-trivial result in this context. Starting with Poincar\'e-AdS, any deformation of the Minkowski boundary metric which preserves asymptotic flatness, and still possesses a bulk extremal horizon extending to boundary null infinity, must possess the same Poincar\'e-AdS near-horizon geometry (\ref{eq:exAdSbulk}).  We will state and argue this statement more precisely in section \ref{sec:application}.

\subsubsection{Non-extension of null infinity to a bulk extremal horizon}

It is instructive to consider two examples where there is a null infinity on the boundary and yet no bulk extremal horizon.

Firstly, consider again a Minkowski boundary, and consider a static bulk corresponding to a thermal state, rather than the vacuum. The bulk is then given by planar Schwarzschild-AdS. 
In this case the boundary metric, being Minkowski, has an extremal null infinity. However the boundary stress tensor corresponds to that of the thermal plasma in the CFT, and is non-vanishing and homogeneous in space. As such, it does not satisfy the fall off requirements of our Result 3 above, and hence null infinity of the boundary Minkowski spacetime cannot extend to an extremal horizon in the bulk. Indeed, planar Schwarzschild-AdS instead of course has a non-extremal horizon. 

Secondly, consider the example of compactifying a Minkowski boundary metric to a Kaluza-Klein space,
\begin{equation}
h_{\text{KK}}= - dt^2 + d\rho^2 + \rho^2  d\Omega_{(d-3)}^{2}  + d\phi^2  \; ,
\end{equation}
where $\phi \sim \phi + L$ parameterises a compact spatial direction. A possible bulk dual is Poincar\'e-AdS with the same identification of the spatial direction; however, this turns the Poincar\'e horizon into a null singularity. Alternatively, another bulk dual is the AdS-soliton~\cite{Horowitz:1998ha}, which has no bulk horizons.  In either case, there is no bulk extremal horizon extending from null infinity.

The fact that the possible bulk duals do not have extremal horizons is consistent with the fact that the boundary metric after identification, no longer has an extremal null infinity. Indeed, conformally compactifying we see,
\be
\bar{h}=\rho^{-2}h_{\text{KK}}  = \frac{-dt^{2} + d\phi^2+ d\rho^2}{\rho^2}+ d\Omega_{(d-3)}^{2}  \; ,
\ee
which is a product of a quotient of Poincar\'e-AdS$_3$ with a $(d-3)$-sphere. In this case null infinity is no longer conformal to a smooth extremal horizon, since this periodic identification of AdS$_3$ renders its Poincar\'e horizon singular. Hence we see explicitly that null infinity is no longer extremal after compactifying a spatial direction of Minkowski. This absence of bulk extremal horizons that meet the boundary is then compatible with our Result 1.

\subsubsection{CFT$_3$ on a static cone and resolved cone}

Consider the case $d=3$ where the CFT$_3$ is defined on a deformation of Minkowski that is no longer asymptotically flat, but instead is a static product of time with a cone,
\begin{eqnarray}
\label{eq:exconeboundary}
h_{\text{cone}} = - dt^2 +  d\rho^2 + \alpha^2 \rho^2 d\phi^2
\end{eqnarray}
for angular coordinate $\phi$ with period $2 \pi$. For $\alpha \ne 1$ the spatial geometry of the boundary is a cone with opening angle $2\pi \alpha$ and is singular at the axis of symmetry $\rho = 0$. 
Null infinity is indeed extremal, as can be seen by changing representative to
\begin{eqnarray}
\label{eq:exconenullinf}
\bar{h} = \rho^{-2}h_{\text{cone}} = \frac{- dt^2 + d\rho^2} {\rho^2}+ \alpha^2 d\phi^2 \; ,
\end{eqnarray}
which is a near-horizon geometry AdS$_2\times S^1$ where the $S^1$ is of radius $2 \pi \alpha$. In particular, we see null infinity is conformal to an extremal horizon at $\rho \to \infty$.

From our results we expect that if null infinity extends to a bulk extremal horizon, then its near-horizon geometry should depend on the  near-horizon geometry conformal to null infinity, and hence depend on $\alpha$. Hence for $\alpha \ne 1$ the bulk horizon's near-horizon geometry will not be that of Poincar\'e-AdS. Indeed this is the case as we now show.

One might imagine the relevant bulk dual is Poincar\'e-AdS written in cylindrical coordinates,
\begin{equation}
\label{eq:exconesing}
g=\frac{-dt^2 + d\rho^2 + \alpha^2 \rho^2 d\phi^2 + dz^{2}}{z^{2}}
\end{equation}
again with angular coordinate $\phi$,
but this is not the case, since the bulk is now singular along the axis of symmetry $\rho = 0$. 

In fact, the general smooth static near-horizon geometry in four dimensional Einstein gravity has been found \cite{Chrusciel:2005pa}. This yields a smooth bulk dual, 
\begin{equation}
\begin{split}
\label{eq:exconebulk}
g &= \psi^2 \left( \frac{-dt^2 +d\rho^2}{\rho^2} \right) +\frac{ d\psi^2}{P(\psi)} + P(\psi) \, \alpha^2 \, d\phi^2  \; ,
\\
P(\psi) &= \left(\psi-\psi_0\right) \left(\psi+ \psi_0  + \frac{\psi_0^2-1}{\psi} \right) \; ,
\end{split}
\end{equation}
where $\psi_0 >1/ \sqrt{3}$ is a constant parameter and the coordinate domain is $\psi > \psi_0$. This geometry has a smooth axis at $\psi=\psi_0$ provided the constant $\alpha$ is fixed in terms of $\psi_0$ by,
\begin{equation}
\alpha = \frac{2\psi_0}{3\psi_0^2-1} \; . \label{coneangle}
\end{equation}
The conformal boundary is at $\psi \to \infty$ and yields the metric \eqref{eq:exconeboundary}. \footnote{
We note that \eqref{eq:exconesing} and \eqref{eq:exconebulk} have different near horizon geometries but the same conformal class of boundary metric, and hence the same boundary null infinity. It is worth observing for \eqref{eq:exconesing} that the near horizon geometry is singular at $\rho = 0$ so that extra data is required for the elliptic near horizon equations at this singularity. It is this singularity and data that results in the difference in the near horizon geometry.
} Clearly the bulk \eqref{eq:exconebulk} has an extremal horizon at $\rho \to \infty$; in fact it is its own near-horizon geometry. Further, we note that the near-horizon geometry is {\it not} that of Poincar\'e-AdS, unless $\psi_0 = 1$ in which case $\alpha=1$. Since (\ref{coneangle}) is a monotonic function of $\psi_0$, we see that the boundary cone opening angle $2\pi {\alpha}$ determines uniquely the near-horizon geometry of this bulk extremal horizon, illustrating our Result 2.

Another nice illustration is to take the boundary metric to be a static resolved cone, asymptotic to \eqref{eq:exconeboundary} but with a smooth axis of symmetry. A particular choice for which we are able to write down a bulk metric, is the boundary metric,
\begin{eqnarray}
\begin{split} \label{eq:resolvedcone}
h_{\text{resolved-cone}} & = - Q(\rho) dt^2 + \frac{\rho^2}{Q(\rho) P(\rho) }d\rho^2 +   P(\rho)\alpha^2 d\phi^2 \; , \\
Q(\rho) &= 1- \frac{\psi_0 (1 - \psi_0^2)}{\rho} \; ,
\end{split}
\end{eqnarray}
where the function $P$ and constant $\alpha$ are as above. This asymptotes to the cone \eqref{eq:exconeboundary} for $\rho \to \infty$, although instead of being singular at the origin the space smoothly caps off at the axis $\rho= \psi_0>0$, so the cone is resolved. Null infinity is again extremal, with near-horizon geometry as $\rho \to \infty$  being precisely that of the cone case \eqref{eq:exconenullinf} above. 

A corresponding smooth static bulk metric to this static resolved cone boundary is an Einstein C-metric\footnote{In fact this is the C-metric used in~\cite{Emparan:1999wa} in the context of 3d brane-world black holes, with $y=\rho^{-1}, x= \psi^{-1}$.},
\begin{equation}
\begin{split}
\label{eq:exconeresolvedbulk}
g &= \frac{1}{(\rho - \psi)^2}\left[  \psi^2 \left( -Q(\rho) dt^2 + \frac{d\rho^2}{Q(\rho)} \right) + \rho^2 \left( \frac{d\psi^2}{P(\psi)} + P(\psi) \alpha^2 d\phi^2 \right)\right] \; ,\end{split}
\end{equation}
in the coordinate domain $\rho>\psi \geq \psi_0$, again for $\psi_0> 1/\sqrt{3}$ with $\alpha$ given by (\ref{coneangle}). As before there is a smooth axis at $\psi= \psi_0$. The conformal boundary is at $ \psi =\rho$ and evaluating $(\frac{\rho}{\psi} -1)^2 g$ there, yields the resolved cone boundary (\ref{eq:resolvedcone}). It is easily shown that this bulk has a smooth extremal horizon at $\rho \to \infty$.\footnote{Introducing new coordinates $r= \rho^{-1}$ and $dv =dt - \frac{d\rho}{Q(\rho)}$, one finds that $r=0$ is a smooth extremal (future) horizon with respect to the Killing field  $\partial /\partial v$.} The near-horizon geometry is found by scaling $(t,\rho) \to (\epsilon^{-1} t, \epsilon^{-1} \rho)$ and sending $\epsilon \to 0$, which yields the general near-horizon geometry \eqref{eq:exconebulk}.

Thus for the same $\alpha$, and hence the same asymptotic conical opening angle, the dual to the cone and to the resolved cone cases indeed both end on an extremal horizon extending from boundary infinity and share the same near-horizon geometry, illustrating Results 1 and 2. Calculating the boundary stress tensor for the boundary metrics in coordinates as in \eqref{eq:exconeboundary} and \eqref{eq:resolvedcone} respectively gives,
\begin{eqnarray}
&&T_{\text{cone}} = c_{\mathrm{eff}} \frac{\psi_0(\psi_0^2-1)}{\rho^3} ( -dt^2+ d\rho^2 - 2\rho^2 \alpha^2 d\phi^2)  \label{coneT}\\
&&T_{\text{resolved-cone}} = c_{\mathrm{eff}} \frac{\psi_0(\psi_0^2-1)}{\rho^3} \left[ - \left( 1 - \frac{3 Q(\rho)}{\rho^2} \right) Q(\rho) dt^2 \nonumber \right. \\  && \quad \qquad \qquad \qquad \qquad \qquad \qquad \left.+ \frac{\rho^2}{Q(\rho) P(\rho)} d\rho^2 + \left( 1 - \frac{3 P(\rho)}{\rho^2} \right)  P(\rho) \alpha^2 d\phi^2 \right]  \; ,
\end{eqnarray}
where $c_{\mathrm{eff}} = 1/16 \pi G_4$. These both decay asymptotically as  $\mathcal{O}(\rho^{-3})$ as $\rho \to \infty$, which is compatible with Result 3. In particular, the leading asymptotics of $T_{\text{resolved-cone}}$ is given by $T_{\text{cone}}$. Also observe that $T_{\text{resolved-cone}}$  is smooth at the axis, as it should be.

Physically we may say that the large scale behaviour of a CFT on a static cone, or static resolved cone with the same asymptotic conical structure, is simply governed by this asymptotic cone geometry. In the bulk this corresponds to the low energy physics living in the highly redshifted region near the extremal horizon whose near-horizon geometry is simply determined by this boundary asymptotic structure.

\section{Null infinity}

\label{sec:null}

In this section we will introduce two types of null infinity. In later sections, we will consider AdS/CFT solutions with boundary spacetimes possessing such null infinities.

\subsection{Extremal null infinity}
\label{sec:Enull}

Let $(\mathcal{M},h)$ be a $d$-dimensional spacetime. Define an unphysical spacetime  $(\overline{\mathcal{M}}, \bar{h})$ which is obtained by attaching a $(d-1)$-dimensional boundary $\mathcal{I}$, where $\bar{h}_{\mu\nu} = \omega^2 h_{\mu\nu}$ for some smooth $\omega>0$ in $\mathcal{M}$ such that $\omega=0$ and $d\omega \neq 0$ on $\mathcal{I}$.  Of course, the unphysical spacetime is only defined up to a strictly positive Weyl transformation $\omega \to e^f \omega$.

Now assume that $\mathcal{I}$ is a  smooth null boundary, so the normal $n_\mu = \partial_\mu \omega$ is null (in the metric $\bar{h}$). We will refer to such spacetimes $(\mathcal{M},h)$ as possessing a {\it smooth null infinity}. This places a constraint on the asymptotic behaviour of the curvature tensor in the physical spacetime. For a general Weyl transformation
\be
R_{\mu\nu}= \bar{R}_{\mu\nu}+ (d-2) \omega^{-1} \bar{\nabla}_\mu \bar{\nabla}_\nu \omega + \bar{h}_{\mu\nu} ( \omega^{-1} \bar{h}^{\rho \sigma} \bar{\nabla}_\rho \bar{\nabla}_\sigma \omega - (d-2) \omega^{-2} \bar{h}^{\rho \sigma} \partial_\rho \omega \partial_\sigma \omega)    \; .
\ee
Multiplying this through by $\omega^2$ and evaluating on $\mathcal{I}$ (i.e. at $\omega=0$) thus implies $n_\mu$ is null if and only if
\be
(\omega^2 R_{\mu\nu}) |_{\mathcal{I}}= 0  \; .  \label{Ricci}
\ee
This of course includes asymptotically flat spacetimes, but also more general behaviour as we discuss below.

We will now assume the existence of a Killing field $\xi^\mu$ of the physical metric whose norm $N= h_{\mu\nu} \xi^\mu \xi^\nu$, and first derivatives, are bounded near $\mathcal{I}$. We may choose $\omega$ such that $\mathcal{L}_\xi \omega=0$. Furthermore, we assume there exists an extension of $\xi^\mu$ to the boundary which leaves the boundary invariant. Then, since $\mathcal{L}_\xi \omega = \xi^\mu n_\mu$ we see that the Killing field is tangent to $\mathcal{I}$. We also deduce that $\mathcal{L}_\xi \bar{h}=0$ on $\overline{\mathcal{M}}$, that is, $\xi^\mu$ is a Killing field of the unphysical spacetime. Also note that if $\xi^\mu$ is hypersurface orthogonal in the physical spacetime it remains so in the unphysical frame.  

Now define the norm in the unphysical spacetime $\bar{N} = \bar{h}_{\mu\nu} \xi^\mu \xi^\nu$. We thus have $\bar{N} = \omega^2 N$.  By assumption $N$ is bounded near $\mathcal{I}$, so it follows that $\xi^\mu$ is null in the unphysical spacetime on $\mathcal{I}$. Since $\xi$ is tangent to $\mathcal{I}$, we see that the Killing field is also normal $\xi^\mu \propto \bar{h}^{\mu\nu}n_\nu$. Therefore, since by assumption $\mathcal{I}$ is a smooth null hypersurface, we deduce that in the unphysical spacetime $\mathcal{I}$ is a Killing horizon of the Killing field $\xi^\mu$. Furthermore, it is easy to see $d \bar{N} = 0$ on $\mathcal{I}$, so that it is in fact a {\it degenerate} Killing horizon.  

We will therefore refer to a null infinity of the above type as an {\it extremal null infinity}. To summarise we have shown the following. \\

\noindent {\bf Lemma}. Consider a spacetime with smooth null infinity $\mathcal{I}$, possessing a one-parameter group of isometries generated by a Killing field $\xi$  tangent to $\mathcal{I}$. If the norm of $\xi$ in the physical metric, and its first derivatives, are bounded near $\mathcal{I}$, there exists a conformal frame where $\mathcal{I}$ is a degenerate Killing horizon of $\xi$. \\

To be more explicit, let us write $\bar{h}$ in coordinates adapted to the extremal horizon $\mathcal{I}$. In Gaussian null coordinates we may write the metric near the horizon as (see e.g.~\cite{Kunduri:2013gce}), 
\be
\bar{h} = r^2 \tilde{\alpha}  dv^2+ 2 dv dr + 2 r \tilde{\beta}_i dv dy^i + \gamma_{ij} dy^i dy^j \; , 
\ee
where the horizon is at $r=0$, the horizon Killing field $\xi = \partial / \partial v$ and the coordinates $(y^i)$ are on a cross-section of $\mathcal{I}$. The functions $\tilde{\alpha}, \tilde{\beta}_i, \gamma_{ij}$ all depend on $(r, y^i)$ and are smooth at $r=0$. 
By construction, the function $\omega=0$ and $d\omega \neq 0$ on $\mathcal{I}$, hence by smoothness it must be of the form $\omega = r e^w$ for some smooth function $w$. Thus, near the horizon, we may use $\omega$ instead of $r$ as a coordinate, which gives
\be
\bar{h} = \omega^2 \alpha  dv^2+ 2\Gamma dv d\omega + 2 \omega \beta_i dv dy^i + \gamma_{ij} dy^i dy^j \; ,  \label{hbar}
\ee
for some functions $\alpha, \beta_i, \gamma_{ij}$ and $\Gamma>0$ which are now smooth functions of $(\omega, y^i)$ at $\omega=0$. In this coordinate system, the near-horizon limit is defined by the rescaling,
\be
(v, \omega, y^i) \to (v/\epsilon, \epsilon \omega, y^i) \; ,  \label{nhlimit}
\ee
for $\epsilon >0$ and letting $\epsilon \to 0$.  Thus, defining a new coordinate,
\be
\rho = \frac{1}{\omega}  \; ,\label{rho}
\ee we see that the physical frame metric is,
\be
h =  \alpha dv^2 -  2 \Gamma dv d\rho + 2 \rho \beta_i dv dy^i + \rho^2 \gamma_{ij} dy^i dy^j \; ,  \label{hphys}
\ee
which has a null infinity at $\rho \to \infty$. Observe that the asymptotic values of the data $\alpha, \beta_i, \gamma_{ij}, \Gamma$ coincide with those of the near-horizon geometry in the unphysical frame.

Let us illustrate the above with a couple of important examples. Consider spacetimes with a static Killing field $\xi =\partial / \partial t$, asymptotic to the direct product of time and a Riemann cone
\be
h_{\text{cone}} =-dt^2+ d\rho^2+ \rho^2 b_{ij}(y) dy^i dy^j \; ,   \label{staticcone}
\ee
where $b_{ij}$ is the metric on the base $B$ of the cone. Observe the norm of $\xi$ is bounded for such spacetimes. This includes (asymptotically) Minkowski spacetimes as a special case. Defining $\omega= \rho^{-1}$ it is clear these spacetimes possess a null infinity at $\omega=0$.\footnote{It is easy to show that  $R_{\mu\nu}= \mathcal{O}(\rho^{-2})$ as $ \rho\to \infty$ so it is clear that (\ref{Ricci}) is satisfied. In fact we have $\omega^{-2} R_{\mu\nu} = \mathcal{O}(1)$ near $\mathcal{I}$.} The metric in the unphysical frame defined by $\omega$ is thus,
\be
\bar{h}= \frac{-dt^2+ d\rho^2}{\rho^2} +  b_{ij}(y) dy^i dy^j \; ,
\ee
which is a near-horizon geometry AdS$_2 \times B$. In particular $\rho \to \infty$ is an extremal horizon, which corresponds to null infinity in the original frame.

As another example consider cosmologies with a spacelike Killing field $\xi = \partial / \partial t$, which at late times are asymptotic to a Milne like universe,
\be
h_\text{Milne} = - d\rho^2+ \rho^2 b_{ij}(y) dy^i dy^j  + dt^2 \; .  \label{milne}
\ee
 Again, such spacetimes possess a null infinity defined by $\omega= \rho^{-1}$ and the norm of the Killing field $\xi$ is bounded. The unphysical metric defined by $\omega$ is thus
\be
\bar{h} = \frac{-d\rho^2+ dt^2}{\rho^2} +  b_{ij}(y) dy^i dy^j \; ,
\ee
which is a near-horizon geometry dS$_2 \times B$. The surface $\rho \to \infty$ is an extremal horizon which corresponds to the late time null infinity of the cosmology.

\subsection{Twisted null infinity}
\label{sec:Tnull}

For some applications it is of interest to relax the notion of extremal null infinity defined above, to allow for other kinds of null infinity. Rather than investigate this possibility in general, we will focus an an explicit class of examples which are closely related to extremal null infinity. 

An important class of spacetimes with extremal null infinities are those given by the static conical spacetimes (\ref{staticcone}). A key property of these spacetimes is the homothety generated by the scaling $(t,\rho) \to (\lambda t, \lambda \rho)$ for constant $\lambda>0$. In fact, the most general static spacetime with such a homothety is given by a `twisted' cone
\be
h_{\text{twisted-cone}} = - a(y)^2 dt^2 + b(y)^2 [d\rho - \rho \, c_i(y)dy^i]^2 + \rho^2 b_{ij}(y) dy^i dy^j  \; ,  \label{twistedcone}
\ee
over a base space $B$ with metric $b_{ij}$.  

It is clear that such spacetimes still possess a null infinity $\mathcal{I}$ defined by $\omega= \rho^{-1}$. The unphysical frame metric is given by 
 \be
\bar{h} =- \frac{a(y)^2 dt^2}{\rho^2} + b(y)^2 \left[\frac{d\rho}{\rho} - c_i(y)dy^i\right]^2+ b_{ij}(y) dy^i dy^j   \; ,  \label{scalegeo}
\ee
which is a generalisation of a static near-horizon geometry. As we discuss in section \ref{sec:twisted} and the Appendix, the null surface at $\rho \to \infty$ in general is not a smooth extremal horizon with respect to $\partial /\partial t$. However, in the special case
\be
c_i = \frac{\partial_i a}{a}- \frac{\partial_i b}{b} \; ,
\ee
it is. This can be seen by redefining $\rho \to (a/b) \rho$ the geometry simplifies to a warped product of AdS$_2$ and $B$,
\be
\bar{h}= b(y)^2 \left( \frac{ -dt^2+ d\rho^2}{\rho^2} \right) + b_{ij}(y) dy^ dy^j \; ,
\ee
which is the general form for a static near-horizon geometry~\cite{Kunduri:2007vf}. 

It is clear that by analytic continuation $a\to i a$ and $b \to ib$ we obtain a twisted version of the Milne like cosmology (\ref{milne}). In this case the above discussion remains valid except for the special case for which they are a warped product of dS$_2$ and $B$.

We will refer to spacetimes with an asymptotic end of the form  (\ref{twistedcone}), or the analytically continued version just mentioned, as possessing a {\it twisted null infinity}.
We will study possible AdS/CFT gravity duals with boundary metrics possessing such a twisted null infinity in section \ref{sec:twisted}.

\section{Extremal horizons in the bulk: The stress tensor as an obstruction}
\label{sec:obstruction}

We will now consider AdS bulk geometries dual to a CFT on $d$-dimensional spacetimes with an extremal null infinity $\mathcal{I}$ as defined in section \ref{sec:Enull}. Since there is a conformal frame on the boundary where null infinity is an extremal Killing horizon, it is natural to ask under what conditions this extends to an extremal horizon in the bulk. As illustrated in our examples in section \ref{sec:examples} this clearly cannot always be the case. We will show that the boundary stress tensor provides a natural obstruction to this extension. We will first explain our general argument and then illustrate this with a simple example.

\subsection{General argument}

Consider AdS bulk spacetimes $(M,g)$ satisfying the Einstein equations (\ref{Einstein}) with a $d$-dimensional a conformal boundary spacetime $(\mathcal{M},h)$. As is well known, the metric in Fefferman-Graham (FG) coordinates adapted to a conformal frame with boundary metric $\bar{h} = \omega^2 h$ is,
\be
g= \frac{dz^2 + \bar{h}_{\mu\nu}(z,x) dx^A dx^B}{z^2}  \; ,  \label{FGcoords}
\ee
where $z=0$ is the conformal boundary. Expanding near $z=0$ and imposing the Einstein equation order by order gives,
\be
\bar{h}_{\mu\nu}(z,x) = \bar{h}_{\mu\nu}(x) + z^2 \bar{h}^{(2)}_{\mu\nu}(x) +\dots +\bar{h}^{(d)}_{\mu\nu}(x) z^d + \bar{a}^{(d)}_{\mu\nu}(x) z^d \log z + \mathcal{O}(z^{d+1})  \; ,  \label{FGexp}
\ee
where only the even powers $z^i$ for $i<d$ appear. The coefficients $\bar{h}^{(i)}$ for $i<d$ and $\bar{a}^{(d)}$ are local covariant expressions determined uniquely by the boundary metric $\bar{h}$.  On the other hand, only the trace and divergence of $\bar{h}^{(d)}$ are determined in terms of $\bar{h}$, so its transverse traceless part $\bar{h}^{(d)}_{TT}$ is free.   The boundary stress-energy tensor $\bar{T}$ fixes this quantity via the relation
\be
\bar{T} = d c_{\text{eff}} \, ( \bar{h}^{(d)} + \bar{X}^{(d)} ) \; ,  \label{stress}
\ee
where $c_{\text{eff}} = 1/(16 \pi G_d)$ and $\bar{X}^{(d)}$ only depends on $\bar{h}$. The quantities $\text{Tr}_{\bar{h}} \bar{h}^{(d)}$,  $\text{div}_{\bar{h}} \bar{h}^{(d)}$, $\bar{a}^{(d)}$ and $\bar{X}^{(d)}$ are dimension dependent; in particular they vanish in all odd dimensions $d$~\cite{deHaro:2000xn}.

Now assume $(\mathcal{M}, h)$ possesses an extremal null infinity $\mathcal{I}$, as defined in section \ref{sec:Enull}. The null infinity $\mathcal{I}$ is an extremal Killing horizon in an unphysical spacetime $(\overline{\mathcal{M}}, \bar{h})$. Therefore, we may write $\bar{h}$ near such a horizon in the form (\ref{hbar}). Any symmetric tensor constructed out of $\bar{h}$, such as $\bar{h}^{(i)}$ for $i<d$ must admit a near-horizon limit (\ref{nhlimit}) and hence have a near-horizon expansion of the same form as the metric itself (see e.g~\cite{Kunduri:2013gce}). It immediately follows that the bulk metric constructed in the FG expansion with all terms of order less than $z^d$ has a smooth extremal Killing horizon extending from null infinity $\mathcal{I}$. 

The question of whether the full bulk spacetime contains an extremal horizon is thus dependant on the behaviour near $\mathcal{I}$ of the next terms in the FG expansion.
As mentioned above,  $\bar{a}^{(d)}$, $\text{Tr}_{\bar{h}} \bar{h}^{(d)}$ and $\text{div}_{\bar{h}} \bar{h}^{(d)}$  are fully determined by $\bar{h}$ and are thus guaranteed to have a near-horizon limit. On the other hand,  the transverse traceless part $\bar{h}^{(d)}_{TT}$ is determined by the boundary stress tensor (\ref{stress}). Now, since $\bar{X}^{(d)}$ is constructed out of $\bar{h}$, it must also admit a near-horizon limit. We deduce that $\bar{h}^{(d)}$ possesses a near-horizon limit if and only if $\bar{T}$ does. In other words, extremal null infinity on the boundary extends to an extremal horizon in the bulk to order $z^d$ in the FG expansion, if and only if $\bar{T}$ possesses a near-horizon expansion of the same form as the boundary metric $\bar{h}$.  Hence, as one would expect, the existence of an extremal horizon in the bulk geometry is dependent on the CFT state.  

We may express this as a condition on the stress-energy tensor $T$ in the physical frame $(\mathcal{M}, h)$ near $\mathcal{I}$. The boundary conformal transformation $h = \omega^{-2} \bar{h}$ induces well-defined transformation properties for the expansion coefficients of the bulk metric. In turn this induces the following conformal transformation of the stress tensor~\cite{Skenderis:2000in},
\be
T= \omega^{d-2} (\bar{T}+ c_d\bar{a}^{(d)})  \;, \label{Ttransformation}
\ee 
where $c_d$ is a constant. The  behaviour near $\mathcal{I}$ of the RHS is thus completely determined by the condition that the bulk metric possesses an extremal horizon extending from null infinity $\mathcal{I}$. This therefore gives a precise decay law for the physical stress tensor near $\mathcal{I}$.

To see this explicitly we first note that any symmetric tensor, such as $\bar{T}+ c_d\bar{a}^{(d)}$, which possesses a near-horizon limit (\ref{nhlimit}) must be of the form,
\be
\bar{T}+ c_d\bar{a}^{(d)} = \omega^2 e dv^2 + 2 f dv d\omega+ 2(h_i \omega dv +l_i d\omega )dy^i + p_{ij} dy^i dy^j  \; ,
\ee
where $e, f, h_i, l_i, p_{ij}$ are smooth functions of $(\omega, y^i)$ at $\omega=0$.  Therefore, using (\ref{Ttransformation}) we deduce that in the physical frame defined by (\ref{hphys}), the stress tensor takes the form,
\be
T = \frac{1}{\rho^d} \left[ e dv^2  - 2f  dv d\rho + 2(h_i \rho dv - l_i d\rho) dy^i  +\rho^2 p_{ij} dy^i dy^j \right]  \; ,  \label{Tphys}
\ee
where $\rho$ is defined by (\ref{rho}).
This shows that the stress tensor must have a fall off $\mathcal{O}(\rho^{-d})$ as one approaches null infinity $\rho \to \infty$ (since the functions $e, f, h_i,l_i, p_{ij}$ are all bounded there). Observe that the physical stress tensor $T \to 0$ near $\mathcal{I}$, a property one would expect for vacuum states. This agrees with the intuition that bulk geometries which end on an extremal horizon are dual to vacuum states in the CFT.

The FG expansion of the bulk metric is determined to all orders by $(\bar{h}, \bar{h}^{(d)})$. However, in general this is merely an asymptotic expansion which does not converge and so does not fully determine the bulk geometry. Therefore, we may only deduce obstructions to the existence of bulk extremal horizons. Namely, null infinity of the boundary extends to a bulk extremal horizon, only if the stress tensor has the decay near $\mathcal{I}$ determined by (\ref{Tphys}).

\subsection{Example: CFT on asymptotically conical static spacetimes}
As an example, suppose our boundary space-time is static and spherically symmetric, with an asymptotically conical boundary metric
\bea
h &&\sim -dt^2+ d\rho^2 + \alpha^2 \rho^2 d\Omega_{d-2}^2   \; ,
\eea
as $\rho \to \infty$, where $\alpha$ is a constant (if $\alpha=1$ it is of course asymptotically Minkowski). Such spacetimes possess a null infinity $\mathcal{I}$ at $\rho \to \infty, t \to \pm \infty$. Letting $\omega=\rho^{-1}$ and $t = v+\rho$, gives the unphysical metric,
\be
\bar{h} \sim  - \omega^2 dv^2+ 2 d\omega dv+ \alpha^2 d\Omega_{d-2}^2  \; ,
\ee
as $\omega \to 0$. In accordance with the general observations made in section (\ref{sec:Enull}), null infinity is an extremal horizon with an AdS$_2 \times S^{d-2}$ near-horizon geometry. 

Assuming that the full bulk geometry inherits the static and spherical symmetry of the boundary, the stress tensor must be static and spherically symmetric so 
\be
T \sim -E(\rho) dt^2 + F(\rho) d\rho^2 + P(\rho) \rho^2 d\Omega_{d-2}^2  \; .
\ee
In the unphysical frame this gives
\be
\bar{T}+c_d \bar{a}^{(d)}  \sim  E\omega^{-d} [-\omega^2 dv^2  + 2 dv d\omega ]+ \omega^{-d} (F-E) d\omega^2 + P \omega^{-d} d\Omega_{d-2}^2 \; ,
\ee
where we have used (\ref{Ttransformation}).
Recall $\bar{a}^{(d)}$ is constructed out of $\bar{h}$ and thus must possess a near-horizon limit. Therefore, the boundary stress tensor $\bar{T}$ possesses a near-horizon limit if and only if
\be
E, F, P = \mathcal{O}(\omega^d)  \; ,
\ee
as $\omega \to 0$.

Converting back to polar coordinates, we deduce that the bulk geometry has an extremal horizon extending from null infinity $\mathcal{I}$ only if the stress tensor satisfies  the fall-offs
\be
E, F, P = \mathcal{O}(\rho^{-d})  \; , \label{stressfalloff}
\ee
as $\rho\to \infty$. We deduce that if the stress tensor in the physical frame does not satisfy the fall off conditions (\ref{stressfalloff}), null infinity of the boundary cannot extend to an extremal horizon in the bulk.

\section{Extremal horizons in asymptotically AdS spacetimes}

\label{sec:extremal}

In this section we will perform a more detailed analysis of asymptotically AdS spacetimes containing extremal Killing horizons which extend to conformal infinity. 
We will first review some general properties of extremal horizons and collect what is known about their near-horizon geometries. Then we will study asymptotically AdS spacetimes containing an extremal horizon which intersects the conformal boundary, which leads to the notion of {\it conformally compact extremal horizons}. 
We will show that for such spacetimes, there is a conformal frame in which the boundary spacetime has an extremal null infinity, as defined in section \ref{sec:Enull} (showing Result 1).
The near-horizon of the bulk extremal horizon is constrained to obey the near-horizon equations which are elliptic p.d.e.s. Solutions are determined by  data which lives at the conformal boundary. We will find that the data that characterises the bulk near-horizon is precisely that which characterises the boundary extremal null infinity (showing Result 2). Finally, we will deduce the asymptotic fall-off of the boundary stress-tensor in the conformal frame where the boundary has a null infinity (showing Result 3).

\subsection{Extremal horizons and their near-horizon geometries}

Here we collect various general properties of extremal horizons and their near-horizon geometries which we will use, see~\cite{Kunduri:2013gce} for more details. 

\subsubsection{General setup}
\label{sec:review}

Consider a $d+1$ dimensional spacetime $(M,g)$ containing a smooth degenerate Killing horizon $\mathcal{N}$ with $(d-1)$-dimensional spatial cross-sections $H$. Let $\xi$ denote the Killing field which is null on the horizon. We will make the following technical assumptions: $\xi$ is hypersurface orthogonal and $H$ is simply connected. 

It can  be shown that the {\it near-horizon geometry} of $\mathcal{N}$ can be written as a warped product~\cite{Kunduri:2007vf},
\begin{equation}
g_{\text{NH}}= \psi(x)^2 d\Sigma_2^2+ \gamma_{ab}(x)dx^adx^b  \; ,  \label{nhstatic}
\end{equation}
where $d\Sigma_2^2$ is a 2d Lorentzian space $M_2$ of constant curvature,  the $(x^a)$ are coordinates on $H$, $\gamma_{ab}(x)$ is the (Riemannian) metric induced on $H$ and $\psi(x)$ is a positive function on $H$. By rescaling $\psi$ we can arrange $M_2$ to have Ricci scalar $2k$ where $k=-1, 0,1$. Then $k=-1,0,1$ correspond to AdS$_2$, $\mathbb{R}^{1,1}$, dS$_2$, respectively. The 2d space written in coordinates adapted to the horizon reads,
\be
d\Sigma^2_2= k \,r^2 dv^2+2dvdr  \; ,  \label{M2}
\ee
where $\mathcal{N}=\left\lbrace r =0 \right\rbrace $ is the horizon and $\xi= \frac{\partial}{\partial v}$.  

It is worth noting that for the AdS$_2$ and dS$_2$ cases one may introduce coordinates adapted to $\xi$ which exhibit its orthogonality. To achieve this set $t = v - k/r$ which gives,
\be
d\Sigma_2^2  = k \left( r^2 dt^2 - \frac{dr^2}{r^2} \right) \; ,  \label{M2static}
\ee
so for $k=-1$ it is AdS$_2$ in Poincar\'e coordinates, whereas for $k=1$ it is dS$_2$ outside the cosmological horizon where $t$ is a spacelike coordinate and $r$ is a timelike one.
On the other hand, for the $\mathbb{R}^{1,1}$ near-horizon geometry the Killing field $\xi$ is everywhere null, so it is not possible to introduce orthogonal coordinates.\footnote{Nevertheless, one may still have a spacetime with such a near-horizon geometry such that $\xi$ timelike/spacelike outside the horizon.}

For such near-horizon geometries the Einstein equations (\ref{Einstein}) are equivalent to the following set of geometrical equations defined on $H$:
\bea
&&R_{ab} = 2 \psi^{-1} \nabla_a \nabla_b \psi - d \gamma_{ab}  \; , \label{Heq1}   \\
&&k = \tfrac{1}{2} \nabla^2 \psi^2 - d \psi^2    \; ,  \label{Heq2}
\eea
where $R_{ab}, \nabla_a$ are the Ricci tensor and metric connection associated to the horizon metric $\gamma_{ab}$. If one assumes $H$ is compact then a global argument reveals that the only solutions are the trivial ones where the function $\psi$ is a constant and the horizon metric is Einstein~\cite{Chrusciel:2005pa}.  We will be interested in solutions with a horizon which possesses a conformal boundary. Thus we will assume $H$ is {\it non-compact}, in which case non-trivial solutions do exist; we will discuss these below.

We will also need the full geometry in a neighbourhood of $\mathcal{N}$, not just its near-horizon limit. To this end, it can be shown that any spacetime containing an extremal horizon, with near-horizon geometry (\ref{nhstatic}), can be written in the form\footnote{This can be shown by working with Gaussian null coordinates and then redefining $r \to \psi(x)^2 r$.}
\be
g = 2 \psi(x)^2 dv \left( dr+ r^2 {k}_a(r,x) dx^a  - \tfrac{1}{2} r^2 {F}(r,x) dv \right) +\gamma_{ab}(r,x) dx^a dx^b \; ,  \label{gnc}
\ee
where all functions are smooth at $r=0$. We should emphasise that hypersurface-orthogonality of $\xi$ imposes non-trivial constraints on the functions $(F,k_a)$, cf.~\cite{Figueras:2011va}. In particular it implies that the 1-form $(k_a dx^a)_{r=0}$ is closed on $H$; this will be important below.
It is easy to see that taking the near-horizon limit $(v,r) \to (v/ \epsilon, \epsilon r)$ and $\epsilon \to 0$ we recover (\ref{nhstatic}) with ${F}|_{r=0} = -k$, so we deduce
\be
F = - k + \mathcal{O}(r) \; ,   \label{F}
\ee
as $r \to 0$. It follows that for horizons with AdS$_2$ ($k=-1$) or dS$_2$ ($k=1$) near-horizon geometry, $\xi$ is timelike or spacelike respectively in a neighbourhood of the horizon.  On the other hand for horizons with the $\mathbb{R}^{1,1}$ ($k=0$) near-horizon geometry there is no such guarantee.

\subsubsection{Near-horizon solutions}
\label{sec:NHG}

In four space-time dimensions ($d=3$) one can in fact determine all solutions to the near-horizon equations (\ref{Heq1})~\cite{Chrusciel:2005pa}.  In particular, if $\psi$ is non-constant  one may use it as a coordinate on $H$ (if $\psi$ is constant, one recovers the trivial solution mentioned above). One then finds the general solution
\be
\label{eq:4dSolution}
\gamma_{ab}dx^a dx^b = \frac{d\psi^2}{P(\psi)} + P(\psi)\alpha^2  d\phi^2  \; ,
\ee
where  $P(\psi) = k+ \beta \psi^{-1} +\psi^2$ and $\alpha, \beta$ are integration constants.  Observe that this metric automatically  possesses a local $U(1)$ isometry. 

By performing a careful global analysis one can show that requiring $H$ to be a complete non-singular manifold, requires the existence of a largest root $\psi_0 >  0$ of $P(\psi)$, so $P'(\psi_0)\geq0$, such that the coordinate domain is $\psi > \psi_0$.\footnote{It is possible to have $\psi_0=0$, although only if $k=0$. The horizon is then locally the hyperbolic plane.} In terms of the parameter $\psi_0$ we may write
\be
P(\psi) = \frac{ (\psi-\psi_0) (\psi^2 + \psi_0 \psi + \psi_0^2+k)}{\psi}  \; ,
\ee
and,
\be
P'(\psi_0) = \frac{3\psi_0^2 +k }{\psi_0} \; .
\ee
If $P'(\psi_0)>0$ there is a conical singularity at $\psi=\psi_0$ which may be removed by identifying $\phi$ with period $2\pi$ and setting
\be
\alpha = \frac{2}{P'(\psi_0)}  \; ,
\ee
in which case there is a smooth axis of the $U(1)$ symmetry.
In this case $H$ has topology $\mathbb{R}^2$ and the horizon is axisymmetric with an axis at $\psi=\psi_0$.  For $k=-1$ this occurs for $\psi_0^2 >1/3$, whereas there is no such constraint for $k=0,1$.  

On the other hand, if $P'(\psi_0)=0$ there is no conical singularity. Hence $\phi$ need not be periodically identified or may be periodically identified with {\it any} period, resulting in horizon topology $\mathbb{R}^2$ and $\mathbb{R} \times S^1$ respectively. In this case there is no axis and instead $\psi \to \psi_0$ is asymptotic to a (quotient of a) hyperbolic cusp. This only occurs for $k=-1$ if $\psi_0^2=1/3$, or for $k=0$ if $\psi_0 \to 0$ (which is just the hyperbolic plane $H^2$).

The AdS$_2$ solution ($k=-1$) generalises a well-known case. For $\psi_0=1$ it is simply the Poincar\'e patch of AdS$_4$ written in coordinates adapted to the Poincar\'e horizon. The explicit coordinate change to Poincar\'e coordinates is:
\be
t = v+ \frac{1}{r}, \qquad z= \frac{1}{\psi r},  \qquad R^2 = \frac{\psi^2-1}{\psi^2 r^2}  \; , \label{Poincarecoords}
\ee
with inverse
\be
r= \frac{1}{\sqrt{z^2+R^2}}, \qquad \psi^{-1} = \frac{z}{\sqrt{z^2+R^2}} \; ,
\ee
which gives,
\be
g= \frac{dz^2 -dt^2+ dR^2+R^2 d\phi^2}{z^2}  \; .
\ee
Hence, the AdS$_2$ near-horizon geometries with $\psi_0^2>1/3$ represent a one-parameter family of generalisations of the Poincar\'e horizon.  Observe these are all axisymmetric.\footnote{The $\mathbb{R}^{1,1}$ solution ($k=0$)  turns out to be the AdS$_4$ soliton written in null coordinates (as opposed to the usual static coordinates).  It may be of interest to investigate the dS$_2$ solution ($k=1$) further.}

In more than four spacetime dimensions $(d>3)$ the full set of solutions to the near-horizon equation (\ref{Heq1}) is not known.  We comment that the Poincar\'e-AdS$_{d+1}$ metric, written in coordinates adapted to the Poincar\'e horizon~\cite{Figueras:2011va, Kunduri:2013gce}, provides a simple example with $k=-1$ and  the horizon metric given by the standard Einstein metric on hyperbolic space $H^{d-1}$,
\be
\gamma_{ab}dx^a dx^b = \frac{d\psi^2}{\psi^2-1} + (\psi^2-1) d\Omega_{d-2}^2  \; .   \label{poincare}
\ee
The coordinate change to Poincar\'e coordinates is the same as in four dimensions above.

In five spacetime dimensions ($d=4$), the general solution with $SO(3)$ symmetry has been found numerically and as in four dimensions turns out to be a one-parameter generalisation of the Poincar\'e horizon~\cite{Kaus:2009cg}; this is the analogue to the general axisymmetric solution in four spacetime dimensions discussed above. Similarly, we expect that in higher dimensions, the general solution with $SO(d-1)$ symmetry is an analogous one-parameter generalisation of the Poincar\'e horizon, although this has yet to be determined.   We should emphasise though, that the space of solutions to (\ref{Heq1}) in higher dimensions $d>3$ is expected to be much larger and more complicated if one relaxes spherical symmetry. 

We now make an important observation. Notice that all the above examples possess non-compact horizons $H$ with a conformal boundary $B$ as $\psi \to \infty$. Namely,  $\psi^{-2}\gamma_{ab} \to \bar{\gamma}_{ab}$ is a non-degenerate metric as $\psi \to \infty$ so that one may define a conformal extension of $H$. We will discuss this from a general point of view in the next section.

Before doing so it is instructive to consider a simple example. For the general four dimensional near-horizon geometry determined by (\ref{eq:4dSolution}), one simply finds $\bar{\gamma}|_{\psi=\infty} = \alpha^2 d\phi^2$ so that the conformal boundary is just a circle (or a line). Interestingly, we see from above that for $k=-1$ there exist two smooth solutions for {\it any} radius $\alpha$, namely an axisymmetric solution with $\psi_0^2>1/3$ and a cusp like solution $\psi_0^2=1/3$. Similarly, for $k=0$ there exist two solutions for any radius $\alpha$, the axisymmetric one with $\psi_0^2>0$ and the hyperbolic cusp geometry $\psi_0=0$. 
However, as discussed above, these two solutions possess a different bulk topology. Nevertheless, this illustrates an important point: for a given conformal boundary, there may exist multiple bulk near-horizon geometries.

Given multiple near-horizon geometries with the same conformal boundary, it is natural to ask how we select the physically relevant ones.  Since these solutions have zero temperature, we will compare their energies. The boundary energy density for the bulk near-horizon geometries dual to the cones (\ref{eq:exconeboundary}) can be read off from the stress tensor (\ref{coneT}), which gives  $\mathcal{E}_{\text{cone}} = c_{\mathrm{eff}} \psi_0(1-\psi_0^2)\rho^{-3}$. For fixed $\rho$, this is in fact maximised by the cusp solution $\psi_0 = 1/ \sqrt{3}$. Therefore, the axisymmetric solution for a fixed $\alpha$, is always energetically preferred to the cusp solution. We deduce that the cusp solution can only be the physically relevant solution if there is some topological reason to discard the axisymmetric solution, e.g. if the theory has fermions with periodic boundary conditions.

\subsection{Conformally compact spacetimes with an extremal horizon}

Consider a spacetime $(M,g)$ which is conformally compact and satisfies the Einstein equations (\ref{Einstein}). This means there exists a conformally related spacetime $(\bar{M}, \bar{g})$ with boundary $\partial M$, such that $\bar{g} = \Omega^2 g$ extends to a non-degenerate metric on $\bar{M}$, for some smooth defining function $\Omega>0$ on $M$ and $\Omega=0, d\Omega \neq 0$ on $\partial M$. As is well known, the Einstein equations imply that the boundary $\partial M$ is timelike with
\be
\bar{g}^{AB} \partial_{A} \Omega \partial_{B} \Omega |_{\partial M} = 1 \; ,  \label{dOmsq}
\ee
for any defining function $\Omega$, and the spacetime $(M,g)$ is asymptotically AdS (in the sense of its curvature). Of course, it is only the conformal class of $\partial M$ which is fixed.

Now suppose the spacetime contains an extremal horizon $\mathcal{N}$ with respect to a hypersurface-orthogonal Killing field $\xi$. As shown above, any such spacetime $(M,g)$ can be written as (\ref{gnc}). We may always choose a defining function which is invariant under the
horizon Killing field, so $\Omega= \Omega(r,x)$. We immediately deduce that $(\bar{M}, \bar{g})$ also contains an extremal horizon at $r=0$ with respect to the same Killing field $\xi$. In particular, the form (\ref{gnc}) implies that $(\bar{M}, \bar{g})$ near the horizon can be written as \footnote{Note that this metric is not strictly in the form \eqref{gnc}. However, it is easy to see that  $r = 0$ is a Killing horizon with respect to $\xi = \partial/\partial v$ with vanishing surface gravity.}
\be
\bar{g} = 2\bar{\psi}(r,x)^2 dv \left( dr+ r^2 {k}_a(r,x) dx^a  - \tfrac{1}{2} r^2 {F}(r,x) dv \right) +\bar{\gamma}_{ab}(r,x) dx^a dx^b \; ,  \
\ee
where the quantities
\be
\bar{\psi} = \Omega \psi, \qquad \bar{\gamma}_{ab}=  \Omega^2 \gamma_{ab} , 
\ee
and $k_a, F$, all extend onto the boundary $\partial M$.   Of course, there is a freedom in the choice of a defining function corresponding to Weyl transformations on $\partial M$. 
Since $\bar{\psi} > 0$ we may use this to select a conformal frame with $\bar{\psi}|_{\partial M}=1$. We will employ this choice below.

It is useful to define the function 
\be
z(x) = \Omega|_{r=0}
\ee 
on the horizon. Observe that in general, the horizon $\mathcal{N}$ and conformal boundary $\partial M$ intersect at the locus $\mathcal{I}$ defined by $(r=0, z(x)=0)$. We will assume there exists a conformal frame $(\overline{\partial M}, \overline{h})$ such that the intersection $\mathcal{I}$ is non-empty and in particular is a smooth hypersurface.  It follows that $\mathcal{I}$ is an extremal horizon in the boundary spacetime $\overline{\partial M}$.  

It turns out to be convenient to use $z$ as a coordinate on cross-sections of the horizon $H$. Restricting (\ref{dOmsq}) to the horizon $r=0$, it easily follows that
\be
(\bar{\gamma}^{ab}\partial_a z \partial_b z )|_{\mathcal{I}} = 1  \; .
\ee
Thus given any representative boundary data $\bar{\gamma}$, by performing a Weyl transformation, we may always find a preferred defining function which satisfies
\be
\bar{\gamma}^{ab}\partial_a z \partial_b z = 1  \;  \label{deffun}
\ee
on $\mathcal{N}$ (at least in a neighbourhood of $\mathcal{I}$). Hence we may introduce Gaussian normal coordinates  $(z,y^i)$ on $H$ adapted to $X= \bar{\nabla} z$ so that the horizon metric reads
\be
\bar{\gamma} = dz^2 + b_{ij}(z,y) dy^i dy^j  \; .  \label{normal}
\ee
In these coordinates the near-horizon geometry (\ref{nhstatic}) of our spacetime $(M,g)$ reads
\be
g_{\text{NH}} =  \frac{ \Gamma(z,y) d\Sigma_2^2+ dz^2 + b_{ij}(z,y) dy^i dy^j}{z^2}  \; ,  \label{FGNH}
\ee
where for notational convenience we have defined the function $\Gamma \equiv \bar{\psi}^2$. As discussed above, we may choose a conformal frame with boundary metric such that $\bar{\psi}|_\mathcal{I}=1$, so $\Gamma(0,y)=1$.  Notice that the near-horizon geometry itself has a conformal boundary at $z=0$ and is automatically in FG coordinates. Its boundary metric is also a near-horizon geometry  $M_2\times B$ where $B$ has metric $b= b_{ij}|_{z=0}dy^i dy^j$.  In the next section we will show how such geometries are determined by specifying just the data $(B,b)$. 

Returning to the full spacetime, we may now exploit the freedom in choosing a conformal frame on the boundary. It turns out to be convenient to work in a conformal frame defined by $z$ itself, so $\bar{g} = z^2 g$. 
\footnote{This is always possible since our assumption that the bulk extremal horizon intersects $\partial M$ on a null hypersurface $\mathcal{I}$ implies $\frac{z}{\Omega}>0$ on $\partial M$, so we may perform a Weyl transformation to set $\Omega \to z$. 
}
Putting things together we deduce that in this frame
\be
g = \frac{2\,\Gamma(z,y)dv \left[ dr+ r^2 (k_z dz+k_i dy^i)  - \tfrac{1}{2} r^2 Fdv \right]  + Adz^2 + C_i dzdy^i +  B_{ij} dy^i dy^j}{z^2} \; , \label{general}
\ee
for some functions $A,C_i, B_{ij}$ suitably well-behaved at $z=0$ -- which satisfy $A|_{r=0}=1, C_{i}|_{r=0}=0$, $B_{ij}|_{r=0} = b_{ij}$ -- and unless otherwise stated all functions depend on all coordinates $(r,z,y^i)$.  Observe that the near-horizon limit of this spacetime coincides with the general class of near-horizon geometries (\ref{FGNH}) as it should. Although in general (\ref{general}) is not in FG coordinates, we may still extract the conformal boundary metric at $z=0$ which is
\be
\bar{h}= 2\, dv \left[ dr+ r^2 k_i(r,y) dy^i  - \tfrac{1}{2} r^2 F(r,y)dv \right] + B_{ij}(r,y) dy^i dy^j   \; .  \label{genbdy}
\ee
This explicitly shows the boundary is also a spacetime with an extremal horizon at $r=0$. From (\ref{F}) we deduce that $F(r,y) = - k +  \mathcal{O}(r)$, so the near-horizon geometry of the boundary spacetime is given by $M_2 \times B$ and so coincides with the boundary of the bulk near-horizon geometry (\ref{FGNH}) (as one would expect). The stress tensor $\bar{T}$ for the full spacetime is harder to extract explicitly since  (\ref{general}) is not in FG coordinates. However, we know that it must admit a near-horizon limit that must coincide with the stress tensor of the near-horizon geometry, which we give in the next section.

\subsection{Conformally compact near-horizon geometries}

\label{sec:horizon}

In the previous section we showed that any conformally compact spacetime which contains an extremal horizon that intersects the conformal boundary, must possess a near-horizon geometry which itself is conformally compact and takes the form (\ref{FGNH}). In this section we study how the Einstein equations constrain such near-horizon geometries.  
Firstly we will argue that the static near-horizon equations are elliptic partial differential equations. Then we will discuss the data that these equations require when the near-horizon geometry has a conformal boundary.

As discussed in section \ref{sec:NHG}, for $d=3$ it is possible to find all solutions to the near-horizon equations (\ref{Heq1}).
For $ d>3$ the general solution is not known. In this case it is convenient to rewrite  the static near-horizon geometry \eqref{nhstatic} as a Kaluza-Klein reduction ansatz,
\begin{equation}
g_{\text{NH}}=  e^{\sqrt{\frac{2 (d-3)}{d-1}} \phi(x)} d\Sigma_2^2 + e^{ - \sqrt{\frac{8}{(d-1)(d-3)}} \phi(x)} h_{ab}(x) dx^a dx^b  \; .
\end{equation}
Then the near-horizon equations (\ref{Heq1}, \ref{Heq2}) take the form,
\bea
&&R^{(h)}_{ab} = \partial_a \phi \partial_b \phi + \frac{2}{d-3} h_{ab} V(\phi) \; , \qquad \qquad  \nabla_{(h)}^2 \phi = V'(\phi) \; ,  \\  &&V(\phi) = - e^{ - \sqrt{\frac{2 (d-1)}{d-3}} \phi} \left( k + \frac{d (d-1)}{2} e^{ \sqrt{\frac{2 (d-3)}{d-1}} \phi} \right) \; .
\eea
These are simply the equations for $(d-1)$-dimensional Einstein gravity in Euclidean signature coupled to a canonical scalar $\phi$ with potential $V(\phi)$. Since there are no two derivative terms of the scalar in the Einstein equation, and none involving the metric in the scalar equation, we may analyse the character of the scalar and Einstein equations separately. The scalar equation is clearly elliptic, the principle part being given by a Laplacian governed by the Riemannian metric $h$. The Einstein equation is also elliptic, provided we fix the coordinate freedom appropriately. The two derivative terms in the form of the Einstein equation above derive solely from the Ricci tensor, whose linearization gives the Lichnerowicz operator. For a Riemannian metric $h$, the Lichnerowicz operator is elliptic taking a harmonic gauge. We deduce that the static near-horizon equations are indeed elliptic p.d.e.s.

We now return to our original form \eqref{nhstatic} of the near-horizon metric. As shown above, for the class of spacetimes under consideration, the near-horizon geometry itself possesses a conformal boundary. In fact, the cross-section $H$ itself is conformally compact with a conformal boundary $B$. \footnote{It is straightforward to define the notion of conformally compact extremal horizons purely from the point of view of the $(d-1)$-dimensional Riemannian manifolds $(H, \gamma_{ab}, \psi)$ which define the near-horizon geometry.} This means that data on the conformal boundary must be provided for the elliptic system of near-horizon equations in order to determine a solution. 

As observed above, the near-horizon geometry (\ref{FGNH}) is automatically in standard FG coordinates (\ref{FGcoords})
where 
\be
 \bar{h}_{\mu\nu}(z,x) dx^\mu dx^\nu= \Gamma(z,y) d\Sigma_2^2+  b_{ij}(z,y) dy^i dy^j   \; .
\ee 
Therefore, we may apply the usual FG expansion to these near-horizon geometries as spacetimes in their own right.  Using (\ref{FGexp}) we deduce analogous expansions for $b_{ij}(z,y)$ and $\Gamma(z,y)$. Using these expansions we find the boundary metric is the direct product $M_2 \times B$,
\be
\bar{h} = d\Sigma_2^2+  b_{ij}(y) dy^i dy^j \; ,  \label{bdynh}
\ee
and the  undetermined coefficient is,
\be
\bar{h}^{(d)} = \Gamma^{(d)}(y) d\Sigma_2^2+  b^{(d)}_{ij}(y) dy^i dy^j   \; .
\ee
Now, using (\ref{stress}) it follows that the stress tensor for the near-horizon geometry must take the form
\be
\bar{T}=  e(y) d\Sigma_2^2+  p_{ij}(y) dy^i dy^j \; ,  \label{stressnh}
\ee
for some functions $e(y), p_{ij}(y)$. In odd dimensions $\bar{X}=0$ and $\bar{h}^{(d)}$ is transverse and traceless, which gives the constraint equations
\be
\text{Tr}_{b} p + 2
e = 0 \; , \qquad
\text{div}_{b} p
=0   \; .
\ee
The even $d$ case is a little more complicated in that both constraint equations pick up extra source terms involving the boundary metric (these can be written down but we will not need them).  

As a simple example, consider Poincar\'e-AdS$_{d+1}$, which can be written as (\ref{FGNH}) where $\Gamma = \left(1+\tfrac{1}{4} z^2 \right)^2$  and $b_{ij} dy^i dy^j = (1-\tfrac{1}{4}z^2)^2 d\Omega_{d-2}^2 $. In this case the boundary data is simply the unit sphere $b =d\Omega_{d-2}^2$ and the undetermined coefficient is given by $\Gamma^{(d)}= \tfrac{1}{16} \delta_{d,4}$, $b^{(d)} = \tfrac{1}{16} \delta_{d,4} d\Omega^2_{d-2}$.

We deduce that this class of near-horizon geometries is determined by the elliptic horizon equations (\ref{Heq1}) on $H$.
Assuming the only boundary of $H$ is the conformal boundary $B$, then these solutions are
 subject only to a choice of boundary metric $b$ on the conformal boundary $B$. Of course, it may be that for given boundary data $(B,b)$ and bulk topology $H$, no solution, multiple solutions or a moduli space of solutions to the elliptic problem 
 exist. These would be distinguished by the stress-tensor data $(e,p)$, just as in the usual case of asymptotically hyperbolic Einstein metrics with prescribed infinity.  

\subsection{Relation between bulk near-horizon geometry and boundary null infinity}

First let us consider a near-horizon geometry of the form (\ref{FGNH}). Write the metric on the 2d space $M_2$ as (\ref{M2}). Then, outside the horizon $r>0$, we can change to a frame with boundary metric $h = r^{-2} \bar{h}$. Defining the coordinate 
\be
\rho = \frac{1}{r},
\ee
we find
\bea
h = k dv^2 - 2 dv d\rho  + \rho^2 b_{ij}(y) dy^i dy^j    \label{nearnull} \; . 
\eea
This spacetime possesses a future null infinity $\mathcal{I}$ at $\rho \to \infty$. Using (\ref{Ttransformation}), we find that in this conformal frame the stress tensor is
\be
T =\rho^{-d}\Big\{ [ e(y)+e^{\text{an}}(y)] (k dv^2 - 2 dv d\rho)  + \rho^2 [p_{ij}(y)+p^{\text{an}}_{ij}(y)] dy^i dy^j \Big\}  \; ,  \label{stressnew}
\ee
where $e^{\text{an}}, p_{ij}^{\text{an}}$ denote the contribution from the anomaly $a^{(d)}$ arising in the transformation (\ref{Ttransformation}).  Interestingly, the anomaly does not actually affect the $\rho$ dependence of the stress tensor.
We see that the data characterising the null infinity, $b_{ij}(y)$, is precisely the same data that specified the boundary conditions for the bulk elliptic near-horizon equations. This establishes our Result 2 for bulk near-horizon spacetimes. 
Shortly, we will extend to the case of bulk spacetimes with extremal horizons with such near-horizon geometries. First however we will examine in more detail the three types of near-horizon geometries.

For the AdS$_2$ case $k=-1$ it is useful to introduce static coordinates (\ref{M2static}), in terms of which 
\bea
h &=& -dt^2 + d\rho^2 +\rho^2 b_{ij}(y) dy^i dy^j \; , \label{NHcone}  \\
T &=& \rho^{-d}\Big\{ [ e(y)+e^{\text{an}}(y)](-dt^2 + d\rho^2) + \rho^2  [p_{ij}(y)+p^{\text{an}}_{ij}(y)] dy^i dy^j \Big\}  \; .
\eea
The boundary is a static Riemann cone over the base $B$ with metric $b$. Observe that null infinity of this static conical spacetime corresponds to the extremal horizon in the original frame. We also see that the stress tensor near $\mathcal{I}$ has a precise $\mathcal{O}(\rho^{-d})$ decay law.

Similarly, for the dS$_2$ case $k=1$ in orthogonal coordinates (\ref{M2static}), we get
\bea
h &=& dt^2 - d\rho^2 +\rho^2 b_{ij}(y) dy^i dy^j \; ,  \label{NHmilne} \\
T &=& \rho^{-d}\Big\{ [ e(y)+e^{\text{an}}(y)](dt^2 - d\rho^2) + \rho^2  [p_{ij}(y)+p^{\text{an}}_{ij}(y)] dy^i dy^j \Big]\} \; .
\eea
In this case, the boundary is the product of a Milne like universe with a flat spacelike direction $t$. Observe that the `late-time' null infinity of this Milne like universe corresponds to the horizon in the original frame, and that the stress tensor must decay at late times as $\mathcal{O}(\rho^{-d})$.

On the other hand, as discussed above, for the $\mathbb{R}^{1,1}$ case $k=0$ it is not possible to introduce coordinates in which the Killing field $\xi$ is manifestly orthogonal. Inspecting (\ref{nearnull}) it is easy to see that $\xi= \partial /\partial v$ is a covariantly constant null vector field. Therefore, the boundary spacetime in this case is a type of pp-wave.

We now wish to generalise the above to a general spacetime (\ref{general}) containing an extremal horizon of these types. As shown above the metric on the conformal boundary is (\ref{genbdy}). 
Hypersurface orthogonality of $\xi$ in particular implies $(k_idy^i)_{r=0}$ is a closed one-form on $B$,  so we can write $(k_idy^i)_{r=0} = d\lambda(y)$ for some function $\lambda(y)$ on $B$ (this is valid globally if $B$ is simply connected). Define a new coordinate
\be
\rho = \frac{1}{r} - \lambda(y) \; .
\ee
Outside the horizon we can change conformal frame to $h= r^{-2} \bar{h}$ and in terms of this new coordinate we find
\be
h =  -2\, dv \left[ d\rho - v_i dy^i  + \tfrac{1}{2} F dv \right] + (\rho+\lambda(y))^2 B_{ij} dy^i dy^j 
\ee
where $F, B_{ij}$ and $v_{i} \equiv k_{i} -k_i|_{r=0}$ are smooth functions at $\rho=\infty$ (this follows from the fact they are smooth functions at $r=0$). Thus as $\rho \to \infty$ we deduce
\bea
F&=& -k+ \mathcal{O}(\rho^{-1}),\qquad B_{ij} = b_{ij}(y)+ \mathcal{O}(\rho^{-1}) \;, \qquad v_i=  \mathcal{O}(\rho^{-1}) \; .
\eea
The leading terms are deduced from their values at $r=0$ (which coincide with those of the near-horizon geometry). Hence in this conformal frame, the boundary spacetime is asymptotic to (\ref{nearnull}) as $\rho \to \infty$ and thus also possesses a future null infinity $\mathcal{I}$ at $\rho = \infty$. This null infinity corresponds to the extremal horizon on the boundary in the original frame. 

Notice in particular that the geometry of null infinity is determined by the boundary near-horizon data $(B,b)$ and vice-versa.   The stress-tensor $T$ for the space-time (\ref{general}) in this frame must asymptote to that derived from the near-horizon geometry (\ref{stressnew}). Therefore we deduce it must have the same leading decay law $\mathcal{O}(\rho^{-d})$. This is consistent with our earlier arguments in section \ref{sec:obstruction}.

To summarise we have shown the following. Firstly, for any asymptotically AdS spacetime which contains an extremal horizon that intersects the conformal boundary, there exists a conformal frame on the boundary with an extremal null infinity. Secondly, the near-horizon geometry itself has a conformal boundary and satisfies the elliptic near-horizon equations with the boundary condition fully determined by the geometry of this null infinity $(B,b)$. Thirdly, the boundary stress tensor in the conformal frame with a null infinity, must have a precise $\mathcal{O}(\rho^{-d})$ decay law.  We have thus established Results 1,2,3 respectively, as stated in section \ref{sec:summary}.

\subsection{Application: CFT on asymptotically flat spacetimes}
\label{sec:application}

Our arguments so far have been rather abstract and general. We will now show how they may be applied to a specific case of interest. Namely, to CFT on any background spacetime which is asymptotically flat. Using our results, we are able to deduce the following general statement regarding the infra-red bulk geometry. \\

\noindent {\bf Conjecture 2}.
Consider a $(d+1)$-dimensional static solution  $(M,g)$ to Einstein's equations (\ref{Einstein}), with a $d$-dimensional conformal boundary $(\partial M, h)$ which is asymptotically Minkowski (\ref{mink}). Suppose $(M,g)$ contains an extremal Killing horizon that extends from the null infinity of $(\partial M, h)$ (and has no other boundaries). Then, the near-horizon geometry must be given by  Poincar\'e-AdS$_{d+1}$ (\ref{eq:exAdSbulk}). Furthermore, the boundary stress tensor must decay  as $\mathcal{O}(\rho^{-d})$.  \\

{\it Argument:} Firstly, we note  the near-horizon geometry must have the conformal boundary AdS$_2 \times S^{d-2}$ with equal radii  (\ref{eq:boundaryMink}). Thus, the spatial sections of the horizon are conformally compact with boundary given by the round unit sphere $S^{d-2}$. Assuming that the $SO(d-1)$ isometry of this boundary extends into the bulk (as is the case for conformally compact Einstein manifolds~\cite{Anderson}), we deduce the near-horizon geometry must possess $SO(d-1)$ symmetry (in fact this assumption is unnecessary for $d=3$, since the general static near-horizon geometry is axisymmetric~\cite{Chrusciel:2005pa}) . Now, for $d=3$ and $d=4$ it is known there is a 1-parameter family of such static near-horizon geometries, with the parameter being the ratio of the radii of AdS$_2$ and $S^{d-2}$ on the boundary~\cite{Kaus:2009cg}. Assuming this is the case in all dimensions, we deduce that fixing the conformal boundary to have equal radii, fixes the near-horizon geometry uniquely.  The fall off of the stress tensor is deduced by conformally rescaling the stress tensor of the bulk near-horizon geometry to the frame in which the boundary metric is that of Minkowski.\\

The above result justifies the choice of bulk boundary condition used in the construction of the black droplet~\cite{Figueras:2011va}.

\section{CFT on backgrounds with a twisted null infinity}

\label{sec:twisted}

In this section we will consider AdS/CFT gravity duals whose boundaries possess a twisted null infinity, as defined in section \ref{sec:Tnull}.  For definiteness, we will consider static spacetimes, although as we mention at the end of this section the cosmological case can be easily obtained by analytic continuation.

Let us suppose the homothety of the boundary metric $h$  (\ref{twistedcone}) is an exact isometry of the bulk. Then, the most general static and scale invariant bulk metric takes the form
\be
g= -  \frac{A(x)^2}{\rho^2} dt^2 + B(x)^2 \left( \frac{d\rho}{\rho} - C_a(x) dx^a \right)^2 + \hat{\gamma}_{ab}(x) dx^a dx^b  \; ,   \label{scale}
\ee
where $\hat{\gamma}_{ab}$ is a (Riemannian) metric on the transverse space $S$ with coordinates $(x^a)$. The scaling isometry in question acts as $(t,\rho) \to (\lambda t, \lambda \rho)$ for constant $\lambda>0$. There is a gauge freedom in our choice of coordinates given by $\rho \to \Gamma(x)^{-1} \rho$, where $\Gamma>0$, which acts as 
\be
A \to \Gamma A, \qquad B \to B, \qquad C_a\to C_a +\Gamma^{-1}\partial_a \Gamma, \qquad \hat{\gamma}_{ab} \to \hat{\gamma}_{ab}  \; . \label{gauge}
\ee 
Thus we see that 
\be
\hat{C}_a  = C_a - \frac{\partial_a A}{A}
\ee
is a gauge invariant quantity.  

Generically, the geometry (\ref{scale}) only possess the static and scaling isometries with 2d orbits. If the transverse space $S$ has Killing fields, it turns out there is an extra gauge freedom. To see this, suppose $K = \partial / \partial x^1$ is a Killing field, so all metric functions are independent of $x^1$. Then, it is clear that there is a gauge freedom in our choice of coordinates $x^1 \to x^1 + k \log \rho$, where $k$ is a constant. This induces certain non-trivial transformation properties for the functions $B, C_a, \hat{\gamma}_{ab}$.

Let us now examine the the null hypersurface $\mathcal{N}$ defined by $\rho \to \infty, t \to \pm \infty$. Clearly the metric (\ref{scale}) is singular on this surface; however,  in some cases it is a mere coordinate singularity. In general, it is easy to show that  if
\be
C_a = \frac{\partial_a A}{A}  - \frac{\partial_a B}{B}  \; ,   \label{regular}
\ee
then $\mathcal{N}$ is in fact a non-singular Killing horizon.
Indeed, performing the gauge transformation (\ref{gauge}) with $\Gamma = B/A$ we get
\be
g=  B(x)^2 \left( \frac{- dt^2 + d\rho^2}{\rho^2} \right) + \hat{\gamma}_{ab}(x) dx^i dx^j \;,  \label{ads2gauge}
\ee
which is general static near-horizon geometry.  Conversely, in the Appendix we show that if $S$ has no continuous isometries, the condition (\ref{regular}) is also necessary for $\mathcal{N}$ to be a non-singular extremal horizon. Thus, {\it generically} the geometries (\ref{scale}) possess a non-singular extremal horizon 
with respect to $\xi$ at $\rho \to \infty$ if and only if (\ref{regular}) is satisfied. If (\ref{regular}) does not hold, then we expect $\rho \to \infty$ is generically a null singularity, although we have not shown this. \footnote{It is easy to see that $\rho=\infty$ is never a curvature singularity.}

In the case of extra transverse Killing fields, the condition (\ref{regular}) is only sufficient for the null hypersurface to be an extremal horizon. 
In the Appendix we consider the special case when $S$ admits a single hypersurface-orthogonal Killing field and derive a condition which generalises (\ref{regular}) that guarantees $\rho = \infty$ is an extremal horizon.

Let us now examine the Einstein equations (\ref{Einstein}) for such scale invariant geometries. We find these are equivalent to a set of geometrical equations defined on the transverse Riemannian manifold $(S, \hat{\gamma}_{ab})$. To see this,  it is convenient to introduce an orthonormal frame
\be
e^0 = \frac{A}{\rho}  dt, \qquad \qquad e^r = B \left( \frac{d\rho}{\rho}-C_a \hat{e}^a \right), \qquad \qquad e^a = \hat{e}^a ,
\ee
where $\hat{e}^a$ are vielbeins for $\hat{\gamma}$. In this frame we find the Einstein equations are equivalent to:
\bea
&&\hat{R}_{ab}  = B^{-1} \hat{\nabla}_a \partial_b B+ \hat{C}_a \hat{C}_b - \hat{\nabla}_{(a} \hat{C}_{b)}  + \tfrac{1}{2}B^2 \hat{\gamma}^{cd} \hat{\nabla}_{[c} \hat{C}_{a]}  \hat{\nabla}_{[d} \hat{C}_{b]}  - d\hat{\gamma}_{ab} \; , \label{ijeq} \\
&&\hat{\nabla}^b \hat{\nabla}_{[b} \hat{C}_{a]}+ \left( -\hat{C}^b+\frac{3\hat{\nabla}^b B}{B} \right) \hat{\nabla}_{[b} \hat{C}_{a]}  - \frac{1}{B^2}\left( \hat{C}_a + \frac{\partial_a B}{B} \right)=0 \; ,\label{rieq} \\
&& \hat{\nabla}^a \hat{C}_a- \hat{C}^a \hat{C}_a + \frac{\hat{C}^a \partial_a B}{B} - \frac{1}{B^2} = -d \; , \\
&& B^2  \hat{\nabla}^{[a} \hat{C}^{b]}  \hat{\nabla}_{[a} \hat{C}_{b]}  - B^{-1} \hat{\nabla}^2 B + \frac{ \hat{C}^a \partial_a B}{B} - \frac{1}{B^2} = -d  \; .
\eea 
These correspond to the $(ab)$, the $(ra)$, the $(00)$ and the $(rr)$ components of the Einstein equations in the above basis, respectively. Observe that they are written entirely in terms of gauge invariant quantities $(B, \hat{C}_a, \hat{\gamma}_{ab})$ defined above, as they should be.  Also note that these equations are not all independent due to the contracted Bianchi identity.

These equations are reminiscent of those for a near-horizon geometry. In fact, imposing the spacetime regularity condition (\ref{regular}), $\hat{C}_a = - B^{-1}\partial_a B$ so $\hat{C}$ is a closed 1-form. It is then easy to see that the equations reduce to
\bea
&&\hat{R}_{ab} = 2 B^{-1} \hat{\nabla}_a \partial_b B - d \hat{\gamma}_{ab}  \\
&&  \tfrac{1}{2} \hat{\nabla}^2 B^2 + \Lambda B^2 =-1 \; ,
\eea
which are precisely those for a  static near-horizon geometry, equations (\ref{Heq1}), (\ref{Heq2}) with $k=-1$ and $B \to  \psi$ and $\hat{\gamma} \to \gamma$.

If one performs the analytic continuation $A\to i A, B \to i B$ in the above, one obtains scale invariant cosmologies which are bulk AdS duals to CFT on boundaries with a null infinity which is a twisted version of the Milne like cosmologies (\ref{milne}).

\subsection{Existence of twisted solutions}
\label{sec:existtwist}

Observe (\ref{rieq}) implies that any solution to the above equations for which $\hat{C}$ is a closed 1-form, must correspond to a regular near-horizon geometry.  It is natural to wonder whether more general solutions exist.  We deduce the necessary and sufficient condition for the existence of twisted solutions is that $\hat{C}$ is not a closed 1-form.  

In fact, we will now prove two non-existence results which provide obstructions to the existence of twisted solutions.  \\

\noindent {\bf  Proposition 1}. Consider a static scale invariant geometry (\ref{scale}), satisfying the Einstein equations (\ref{Einstein}), with a transverse space $S$ which is compact with no boundary.  The geometry must be a warped product of AdS$_2$ with $S$ and possesses $SO(2,1)$-symmetry. \\

\noindent {\it Proof.} Define the vector $X_a = \hat{C}_a +B^{-1} \partial_a B$, in terms of which (\ref{rieq}) can be written as
\be
\hat{\nabla}^b( B^4 \hat{\nabla}_{[b} X_{a]} ) - X^b B^4 \hat{\nabla}_{[b} X_{a]}   - B^2 X_a=0   \; .
\ee
By contracting this equation with $X^b$, one can show that
\be
\hat{\nabla}^b( X^a B^4 \hat{\nabla}_{[b} X_{a]} )  = B^4 \hat{\nabla}^{[b} X^{a]}   \hat{\nabla}_{[b} X_{a]} +B^2 X^a X_a   \; .   \label{identity}
\ee
Since $S$ is compact with no boundary, by integrating this over $S$ we immediately deduce that $X_a \equiv 0$ everywhere, so (\ref{regular}) is satisfied.  One may then perform a gauge transformation to get (\ref{ads2gauge}), thus establishing the claim. \\

Hence for compact $S$ there are no more general static scale invariant geometries than the standard static near-horizon geometries.  It is interesting to note that for scale invariant cosmologies, the above argument fails (since $B \to iB$ spoils positivity). Thus there may be the possibility of scale invariant geometries with compact $S$ more general that the standard warped dS$_2$ near-horizon geometries. We will not pursue this here, but this may be worth investigating further.

Of course, we are interested in the case where $S$ is a complete manifold with a conformal boundary (so non-compact). Consider a 4-dimensional static bulk solution so $d=3$.
As we have reviewed in section \ref{sec:NHG} in this case static near-horizon geometries must have an extra $U(1)$ isometry and the general solution is that in equation  \eqref{eq:exconebulk}. 
We might wonder if it is possible to similarly find twisted solutions in this $U(1)$ symmetric class. In fact it is not. \\

\noindent {\bf Proposition 2}. Consider a four-dimensional static, scale invariant, $U(1)$-symmetric spacetime of the form (\ref{scale}), satisfying the Einstein equations (\ref{Einstein}). Then the spacetime must be a warped product of AdS$_2$ with $SO(2,1)\times U(1)$ symmetry. \\

\noindent{\it Proof}.  We begin by writing a general form for a static, scale invariant, $U(1)$ symmetric bulk metric as
\begin{eqnarray}
g = - \frac{A(\psi)^2}{\rho^2} dt^2 +  \frac{B(\psi)^2}{\rho^2} d\rho^2 + \frac{d\psi^2}{P(\psi)} + P(\psi) d\phi^2 + \frac{2 F(\psi)}{\rho} d\rho d\phi   \; .
\end{eqnarray}
In the above we have made the following gauge choices. Firstly, on the transverse space $S$, we have introduced coordinates $(\psi, \phi)$ which are adapted to the $U(1)$ symmetry such that $\partial / \partial \phi$ is the Killing field and $g_{\phi\phi} = g_{\psi\psi}^{-1}$ and $g_{\psi \phi}=0$. Secondly, we have exploited the gauge freedom (\ref{gauge}) to set $g_{\rho \psi}=0$.  Notice that positive definiteness of the spatial metric in particular requires $B^2P-F^2>0$. Then one finds that the $(\psi\phi)$ and  $(r \psi)$ components of the Einstein equations (\ref{Einstein}) can be  integrated to yield that $F = k P$ and $B^2  = c A^2 + k^2 P$ for constants of integration $k,c$, so positive definiteness requires $c>0$. Thus the solution must take the form,
\begin{eqnarray}
g &=& \frac{A(\psi)^2}{\rho^2} \left( - dt^2 + c \, d\rho^2 \right) + \frac{d\psi^2}{P(\psi)} + P(\psi) \left( d\phi + \frac{k}{\rho} d\rho \right)^2  \; .
\end{eqnarray}
Redefining $t \to \sqrt{c} t$ and $\phi \to  \phi - k \log \rho$ we recognise this is simply a static near-horizon geometry. Hence the only static,  $U(1)$ symmetric, twisted solutions are in fact the untwisted near-horizon geometries, as claimed.\footnote{Observe that this example violates condition (\ref{regular}), but still possesses a smooth degenerate horizon at $\rho =\infty$ due to the presence of the $U(1)$ isometry.}  \\

 From the AdS-CFT perspective the above result is related to the fact that there is no deformation of the boundary metric dual to the near-horizon \eqref{eq:exconebulk} with $U(1)$ symmetry that introduces a twist. Recall that the boundary metric is represented by the static cone,
\begin{eqnarray}
h_{\text{cone}} = -dt^2 + d\rho^2 + \rho^2 \alpha^2 d\phi^2 \; ,
\end{eqnarray}
where $\phi$ is an angle and the constant $\alpha$ sets the spatial cone opening angle. Now, the only possible form of a twisted cone metric, preserving the static, scaling and $U(1)$ isometries, is
\begin{equation}
\begin{split}
h &= - dt^2 + d\rho^2 + \rho^2 \alpha^2 d\phi^2 + 2 \beta \rho d\rho d\phi   \; ,
\end{split}
\end{equation}
where $\alpha, \beta$ are constants (we may always rescale the coordinates to set $h_{tt} = -1$ and $h_{\rho\rho}=1$). However, we may rewrite this as,
\begin{equation}
\begin{split}
h_{} &= - dt^2 + \left( 1 - \frac{\beta^2}{\alpha^2} \right) d\rho^2 + \rho^2 \alpha^2 \left( d\phi + \frac{\beta}{\alpha^2} \frac{d\rho}{\rho} \right)^2 \; , 
\end{split}
\end{equation}
so redefining $\phi' = \phi + \frac{\beta}{\alpha^2} \log \rho$, and $\rho' =  \rho \sqrt{1 - \beta^2/\alpha^2}$ we see there is in fact no twist. Thus the boundary metric is equivalent to a static cone, and hence we expect the bulk dual to have an extremal horizon provided there is no obstruction from the stress tensor.

Thus in order to introduce a twist in this static 4-dimensional case we see that we must break the $U(1)$ symmetry, and look for solutions to the resulting 2-dimensional p.d.e. system.  We have not found such solutions, although we expect them to exist. Specifically, if one specifies a boundary metric that is twisted of the form \eqref{twistedcone}, we expect a twisted bulk solution to exist, although for the reasons above it may be challenging to find them.  In the next section we give evidence towards their existence.

Before moving on, we note that the above two results are reminiscent of the near-horizon symmetry enhancement theorems~\cite{Kunduri:2007vf, Figueras:2008qh, Lucietti:2012sa}. In that case a general near-horizon geometry only possesses time-translation plus scaling-symmetry and under certain assumptions the Einstein equations imply this is enhanced to $SO(2,1)$. Here we have instead started from a general static scale invariant geometry and shown that under some assumptions the Einstein equations imply an enhancement of symmetry to $SO(2,1)$ (resulting in a non-singular geometry).

\subsection{Example: A twisted deformation of Poincar\'e-AdS}
\label{sec:twistedpert}

Unfortunately we have not been able to solve the Einstein equations for  non-trivial twist. This is not surprising since the near-horizon equations themselves are already difficult to solve.  

As discussed above, we expect that four-dimensional static, scale-invariant, non-axisymmetric twisted solutions to exist. In order to provide evidence for this claim we now show that one may introduce a small twist to Poincar\'e-AdS$_4$ and solve the bulk equations to linear order.

We write the static metric as,
\begin{eqnarray}
g &=&  \frac{\psi^2}{\rho^2} \left( - dt^2 + d\rho^2 \right) + \frac{1}{\psi^2 - 1} d\psi^2 + \left( \psi^2 - 1 \right) d\phi^2 \nonumber \\
&& \quad + \epsilon \left( - \frac{1}{\psi^2 - 1} P d\psi^2 + \left( \psi^2 - 1 \right) P d\phi^2 + \frac{2}{\rho} J d\rho d\psi + \frac{2}{\rho} F d\rho d\phi + 2 H d\psi d\phi \right) \; ,
\end{eqnarray}
where we will work to linear order in $\epsilon$, and $P$, $J$, $F$ and $H$ are functions of $\psi$ and $\phi$. We decompose the perturbation in Fourier modes on the circle with coordinate $\phi$ as,
\begin{eqnarray}
P(\psi, \phi) &= & p(\psi) \cos{n \phi} \; , \quad  J(\psi, \phi) = j(\psi) \cos{n \phi} \; , \nonumber \\
F(\psi, \phi) &= & f(\psi) \sin{n \phi} \; , \quad  H(\psi, \phi) = h(\psi) \sin{n \phi}  \; ,
\end{eqnarray}
for integer $n$. We have chosen a particular gauge for the perturbation. We note that the most general static perturbation that is invariant under the scaling $(t, \rho) \to (\lambda t, \lambda \rho)$, with this harmonic dependence on $\phi$, can be brought into this form by an infinitesimal diffeomorphism generated by the vector field, 
\begin{equation}
\begin{split}
\chi = \rho \, g_1(\psi) \cos{n \phi} \frac{\partial}{\partial \rho} + g_2(\psi) \cos{n \phi} \frac{\partial}{\partial \psi} + g_3(\psi) \sin{n \phi} \frac{\partial}{\partial \phi}   \; ,
\end{split}
\end{equation}
for some appropriately chosen functions $g_{1,2,3}$ which are determined purely algebraically. The linear perturbation equations then have general solution where $p(\psi)$ obeys a second order o.d.e. with two solutions,
\begin{eqnarray}
p(\psi) = \frac{1}{\psi \left( \psi^2 - 1 \right)^{\frac{n+2}{2}}}  \left[  c_1 \left( \psi + 1 \right)^n \left( n \psi + 1 \right) + c_2 \left( \psi - 1 \right)^n \left( n \psi - 1 \right) \right] \; ,
\end{eqnarray}
for constants $c_1$ and $c_2$ with the other functions determined in terms of $p(\psi)$ and its first derivative. The perturbation parameterized by $c_1$ is singular at the $U(1)$ symmetry axis $\psi = 1$. Remarkably, the perturbation parameterized by $c_2$ is actually regular at the axis for all integer $n \geq 1$. For this regular perturbation, setting $c_1 = 0$ and $c_2 = 1$, the metric functions take the form,
\begin{eqnarray}
p(\psi) &=& h(\psi) =  \frac{ \left( \psi - 1 \right)^{\frac{n}{2} - 1} }{ \left( \psi + 1 \right)^{\frac{n}{2} + 1} } \left( n - \frac{1}{\psi} \right) \; ,\nonumber \\
 j(\psi) &=& - \frac{n \psi}{ n  - \frac{1}{\psi}} p(\psi) \; , \qquad f(\psi) = \frac{\left( \psi^2 - 1 \right)\left( \psi^2 + n \psi - 1\right)}{ n - \frac{1}{\psi} }  p(\psi)  \; .
\end{eqnarray}
One can explicitly check this perturbation is indeed smooth on the axis $\psi=1$, for every integer $n \geq 1$, by transforming to Cartesian coordinates $(x,y)$, so that $\psi^2 -1= x^2 + y^2$ and $\tan{\phi} = y/x$. The calculation is a little involved, although straightforward, so we omit the details. 

This bulk perturbation then generates a boundary perturbation,
\begin{equation}
\begin{split}
h= - dt^2 + d\rho^2 + \rho^2 d\phi^2 + 2 \epsilon \, \rho \sin{n \phi} \,d\rho d\phi  \; ,
\end{split}
\end{equation}
which precisely introduces a twist term onto the background Minkowski boundary metric, provided that $n \ne 0$. Note that unlike the $U(1)$ symmetric case discussed above, in this case the twist is genuine and cannot be removed by coordinate transformations. Thus by superposing the harmonic modes we see that there is a single regular bulk solution for the boundary metric,
\be
h = - dt^2 + d\rho^2 + \rho^2 d\phi^2 + 2 \epsilon \, c(\phi) \, \rho  \,d\rho d\phi  \; ,
\ee
where $c(\phi)$ is any periodic function. This establishes our above claim.

We note that for the special case of a perturbation just composed of the $n = 0$ mode, this perturbation does not deform the boundary. Although the above perturbation in this case looks singular on the axis, in fact it is gauge equivalent to the bulk perturbation tangent to the C-metric moduli space, so that $\psi_0 - 1 = \epsilon / 2$ for the C-metric in equation \eqref{eq:exconeresolvedbulk}, to linear order in $\epsilon$.

\section*{Acknowledgements}
We would like to thank Centro de Ciencias de Benasque Pedro Pascual for hospitality, where this work was initiated. We would also like to thank Pau Figueras and Hari Kunduri for useful discussions. JL is supported by an EPSRC Career Acceleration Fellowship. AH is supported by an STFC studentship.

\appendix

\section{On the regularity of static scale invariant geometries}

Rewrite (\ref{scale}) in terms of $r = \rho^{-1}$, so
\be
g= -  r^2A(x)^2 dt^2 + B(x)^2 \left( \frac{dr}{r} + C_a(x) dx^a \right)^2 + \hat{\gamma}_{ab}(x) dx^a dx^b  \; .   \label{scale2}
\ee
Consider the null hypersurface $\mathcal{N}$ defined by $r=0$. In the above coordinates the metric is clearly singular on this surface. We wish to find general conditions on $A,B,C_a$ which are necessary and sufficient for $\mathcal{N}$ to be a non-singular extremal Killing horizon with respect to the Killing field $\xi = \frac{\partial}{\partial t}$.   

To proceed we assume $\mathcal{N}$ is a smooth (in fact $C^2$) extremal Killing horizon with respect to the Killing field $\xi$. It is then well known that a coordinate system, called Gaussian null coordinates, may be constructed in the neighbourhood of $\mathcal{N}$, as follows (see~\cite{Kunduri:2013gce} for more details). Let $S$ be a cross-section of $\mathcal{N}$ and $\tilde{x}^a$ coordinates on $S$. Starting from some point on $S$ consider the point on $\mathcal{N}$ a parameter value $v$ along the flow of $\xi$ and assign it coordinates $(v,\tilde{x}^a)$. Consider null geodesic vectors $U$ which `shoot-out' from a point on $S$ transverse to $\mathcal{N}$ and satisfy
\be
U \cdot \xi = 1, \qquad U \cdot X = 0  \label{Udef}
\ee
on $\mathcal{N}$, where $X$ is any tangent vector on $S$.  By the geodesic property of $U$, these conditions remain true in a neighbourhood of $\mathcal{N}$. Assign coordinates $(v, \lambda, \tilde{x}^a)$ to the point reached a parameter value $\lambda$ along such geodesics. This gives a coordinate system $(v, \lambda, \tilde{x}^a)$ valid in a neighbourhood of $\mathcal{N}$ called Gaussian null coordinates.  Observe that in these coordinates
\be
\xi = \frac{\partial}{\partial v}, \qquad U = \frac{\partial}{\partial \lambda} \; ,
\ee
and $\partial / \partial \tilde{x}^a$ are tangent to $S$. We now apply this construction to the above metric (\ref{scale2}). The analysis depends on whether (\ref{scale2}) possesses continuous isometries transverse to the orbits of the static and scale isometries. 

If $S$ possesses no Killing fields, the  $(x^a)$ may be taken as coordinates on the cross-section $S$,  so a basis for tangent vectors to $S$ is given by
\be
X_a = \frac{\partial}{\partial {x}^a} - C_a r \frac{\partial}{\partial r}  \; .  \label{Xi}
\ee
(These are the dual vectors to $dx^a$). Parameterising the geodesics $U = (\dot{t}, \dot{r}, \dot{x}^a)$, it is then easy to show that the conditions (\ref{Udef}) reduce to
\be
\dot{t} = - \frac{1}{A^2r^2}\; ,  \qquad \dot{x}^a=0  \; ,
\ee
respectively.  It then follows that the null constraint $U \cdot U=0$ reduces to
\be
\dot{r} = \frac{1}{AB}  \; .
\ee
These geodesic equations are easily solved to give
\bea
t (\lambda)= v + \frac{B(\tilde{x})^2}{\lambda}, \qquad r(\lambda)=\frac{1}{A(\tilde{x})B(\tilde{x})} \lambda, \qquad x^a(\lambda) = \tilde{x}^a   \; ,\label{geodesics}
\eea
where $(v, \tilde{x}^a)$ are constants and we fixed the integration constant for $r(\lambda) $ so that $\lambda=0$ corresponds to $\mathcal{N}$. Therefore, we have obtained a family of null geodesics which shoot-out from $\mathcal{N}$ parameterised by $(v, \tilde{x}^a)$.  Thus, according to the above construction, we have obtained a Gaussian null coordinate system $(v, \lambda, \tilde{x}^a)$ valid in a neighbourhood of $\mathcal{N}$ which is related to our original coordinates by (\ref{geodesics}).

In fact it is convenient to use a different affine parameter for $U$, namely $\tilde{\lambda} = B^{-2} \lambda$, in which case
\be
t = v+ \frac{1}{\tilde{\lambda}}, \qquad r = \frac{B(\tilde{x})}{A(\tilde{x})}  \tilde{\lambda}, \qquad x^a = \tilde{x}^a \; .
\ee
Using this coordinate change (\ref{scale2}) becomes
\bea
g &=&  B(\tilde{x})^2 ( - \tilde{\lambda}^2 dv^2 + 2 dv d\tilde{\lambda})  + 2 B^2\left( C_a+ \frac{\partial_a B}{B} - \frac{\partial_a A}{A} \right) \frac{ d\tilde{\lambda} d\tilde{x}^a}{\tilde{\lambda}} \nonumber \\ &+& \left[ \hat{\gamma}_{ab}(\tilde{x}) + \left( C_a+ \frac{\partial_a B}{B} - \frac{\partial_a A}{A} \right)\left( C_b+ \frac{\partial_b B}{B} - \frac{\partial_b A}{A} \right) \right] d\tilde{x}^a d\tilde{x}^b  \; .   \label{singular}
\eea
We deduce that this metric is non-singular at $\tilde{\lambda}=0$ if and only if
\be
C_a =  \frac{\partial_a A}{A} -  \frac{\partial_a B}{B}  \label{Ci}
\ee
is satisfied. In this case, the resulting metric is 
\be
g =  B(\tilde{x})^2 ( - \tilde{\lambda}^2 dv^2 + 2 dv d\tilde{\lambda}) + \hat{\gamma}_{ab}(\tilde{x}) d\tilde{x}^a d \tilde{x}^b  \; ,
\ee
which is a warped product of AdS$_2$ and the cross-section $(S, \hat{\gamma}_{ab})$ and is the general form for the near-horizon geometry of a static extremal Killing horizon~\cite{Kunduri:2007vf}.

If $S$ has Killing fields the situation is a little more complicated. Suppose $S$ has a single Killing field $K$ and choose coordinates adapted to this so $K=\partial/ \partial x^1$. Then, as observed in section \ref{sec:twisted}, the diffeomorphism $x^1 \to x^1 + k \log r$ for constant $k$ is a gauge freedom. This means the cross-section $S$ may not necessarily be located at finite value of the coordinate $x^1$, so the above analysis needs to be modified. Rather than repeat the above analysis in general, we will simply show that (\ref{Ci}) is not necessary for $r=0$ to be a non-singular extremal horizon, as follows.

For simplicity, suppose $K$ is hypersurface-orthogonal. Then, in static coordinates (\ref{singular}) can be written as
\bea
g &=&  B(\tilde{x})^2 \left[ - \tilde{\lambda}^2 dt^2 + \left(1- \frac{B^2 C_1^2}{\gamma_{11}} \right)\frac{d\tilde{\lambda}^2}{\tilde{\lambda}^2} \right] + \gamma_{11} \left( d\tilde{x}^1+ \frac{B^2 C_1}{\gamma_{11}} \frac{d\tilde{\lambda}}{\tilde{\lambda}}\right)^2  \\ &+& 2 B^2\left( C_I+ \frac{\partial_I B}{B} - \frac{\partial_I A}{A} \right) \frac{ d\tilde{\lambda} d\tilde{x}^I}{\tilde{\lambda}}+ {\gamma}_{IJ}(\tilde{x}) d\tilde{x}^I d\tilde{x}^J  \; .  \nonumber
\eea
where $\gamma_{ab}=\hat{\gamma}_{ab}+\left( C_a+ \frac{\partial_a B}{B} - \frac{\partial_a A}{A} \right) \left( C_b+ \frac{\partial_b B}{B} - \frac{\partial_b A}{A} \right)$ and $I,J=2,\dots, d-1$. We deduce that the surface $\tilde{\lambda}=0$ is non-singular if (\ref{Ci}) is satisfied for $a=2, \dots d-1$ and 
\be
\frac{B^2 C_1}{\gamma_{11}} = k \; ,
\ee
for some constant $k$, and $C_1$ is constant. Observe that for $k \neq 0$ this latter condition generalises (\ref{Ci}) for $a=1$. In this case the geometry is 
\be
g = B(\tilde{x})^2 \left[- \tilde{\lambda}^2 dt^2 + \left(1-k \, C_1 \right)\frac{d\tilde{\lambda}^2}{\tilde{\lambda}^2} \right] + \gamma_{11} dy^2 + \hat{\gamma}_{IJ} d\tilde{x}^I d \tilde{x}^J  \; ,
\ee
where $y= \tilde{x}^1+ k \log \lambda$, so we also require $k<1$, resulting in a warped product of AdS$_2$. Notice that coordinates on the cross-section $S$ in this case are given by $(y, \tilde{x}^I)$, so, as alluded to above, in the original coordinates $S$ corresponds to $x^1 \to \infty$ with $x^I$ finite.

\end{document}